%% file: Josephson_diode_PRB.tex
\documentclass[aps,prb,10pt,twocolumn,a4paper,superscriptaddress,nofootinbib,longbibliography]{revtex4-2}

\usepackage[colorlinks = true,
            linkcolor = blue,
            urlcolor  = blue,
            citecolor = blue,
            anchorcolor = blue,
            unicode]{hyperref}

\usepackage{graphicx}
\usepackage{amsmath}
\usepackage{amssymb}
\usepackage{amsfonts}
\usepackage{color}

\DeclareMathOperator{\sgn}{sgn}

\DeclareMathOperator{\maxmin}{max(min)}

\renewcommand{\Re}{\mathop\mathrm{Re}\nolimits}
\renewcommand{\Im}{\mathop\mathrm{Im}\nolimits}

\begin{document}

\title{Asymmetric higher-harmonic SQUID as a Josephson diode}

\author{Ya.~V.\ Fominov}
\affiliation{L.~D.\ Landau Institute for Theoretical Physics RAS, 142432 Chernogolovka, Russia}
\affiliation{Laboratory for Condensed Matter Physics, HSE University, 101000 Moscow, Russia}

\author{D.~S.\ Mikhailov}
\affiliation{Moscow Institute of Physics and Technology, 141700 Dolgoprudny, Russia}

\begin{abstract}
We theoretically investigate asymmetric two-junction SQUIDs with different current-phase relations in the two Josephson junctions, involving higher Josephson harmonics.
Our main focus is on the ``minimal model'' with one junction in the SQUID loop possessing the sinusoidal current-phase relation and the other one featuring additional second harmonic.
The current-voltage characteristic (CVC) turns out to be asymmetric, $I(-V) \neq -I(V)$.
The asymmetry is due to the presence of the second harmonic and depends on the magnetic flux through the interferometer loop, vanishing only at special values of the flux such as integer or half-integer in the units of the flux quantum.
The system thus demonstrates the flux-tunable Josephson diode effect (JDE), the simplest manifestations of which is the direction dependence of the critical current.
We analyze asymmetry of the overall $I(V)$ shape both in the absence and in the presence of external ac irradiation.
In the voltage-source case of external signal, the CVC demonstrates the Shapiro spikes. The integer spikes are asymmetric (manifestation of the JDE) while the half-integer spikes remain symmetric.
In the current-source case, the CVC demonstrates the Shapiro steps.
The JDE manifests itself in asymmetry of the overall CVC shape, including integer and half-integer steps.
\end{abstract}

\date{9 December 2023}

\maketitle

\tableofcontents

\section{Introduction}
\label{sec:intro}

The diode effect in electronic charge transport is fundamentally important from the scientific point of view, and finds numerous applications in devices like rectifiers, filters, and current converters \cite{MalvinoBook}.
At present, its superconducting version, the superconducting diode effect (SDE) is being actively studied both theoretically and experimentally.
This topic is particularly interesting because the SDE (nonreciprocity of supercurrent) can arise in various physical systems due to different mechanisms.

First, the SDE can arise in bulk (extended) superconducting systems \cite{Silaev2014,Wakatsuki2017,Yasuda2019,Daido2022.PhysRevLett.128.037001,Yuan2022,Ilic2022.PhysRevLett.128.177001,He2022,
Karabassov2022arXiv,Kokkeler2022arXiv}.
The necessary ingredients are usually broken time-reversal and inversion symmetries, which can be realized, e.g., due to magnetic field (or exchange field in ferromagnets) and spin-orbit coupling. Topological materials and materials without a center of inversion can be advantageous for achieving the necessary requirements; breaking of the inversion symmetry can also be achieved due to artificial structuring \cite{Ando2020,Lyu2021}.
The SDE can also be realized due to vortices pinned or controlled by various types of asymmetric forces (asymmetric configurations of magnetic field, asymmetric traps, or asymmetric surface barriers) \cite{Majer2003.PhysRevLett.90.056802,Villegas2003,Vodolazov2005.PhysRevB.72.064509,deSouzaSilva2006,Morelle2006,Suri2022}; in the context of vortex motion the SDE is often referred to as the ratchet effect.
(At the same time, to avoid misunderstanding, we note that the physics we are discussing now is different from the \emph{stochastic} ratchet effect that arises in nonequilibrium dynamics of nonlinear systems with asymmetric potential \cite{Julicher1997.RevModPhys.69.1269,Doering1998}. In those systems, a directed transport can arise due to noise even in the absence of a bias. We do not consider noise effects in this paper.)

Second, similar physical mechanisms \cite{Yokoyama2014.PhysRevB.89.195407,Chen2018.PhysRevB.98.075430,Kopasov2021.PhysRevB.103.144520,Baumgartner2022,Halterman2022.PhysRevB.105.104508,Pal2022,
Zhang2021arXiv,Davydova2022arXiv}
or certain charging effects \cite{Hu2007.PhysRevLett.99.067004,Misaki2021.PhysRevB.103.245302,Wu2022}
can lead to the SDE in Josephson junctions (JJs); in this context it can be called the Josephson diode effect (JDE).
This bring the rich physics of the Josephson effect \cite{BaroneBook,LikharevBook} into play. Asymmetry of the Josephson effect with respect to the current direction implies realization of the JDE.

\begin{figure}[t]
 \includegraphics[width=0.45\columnwidth]{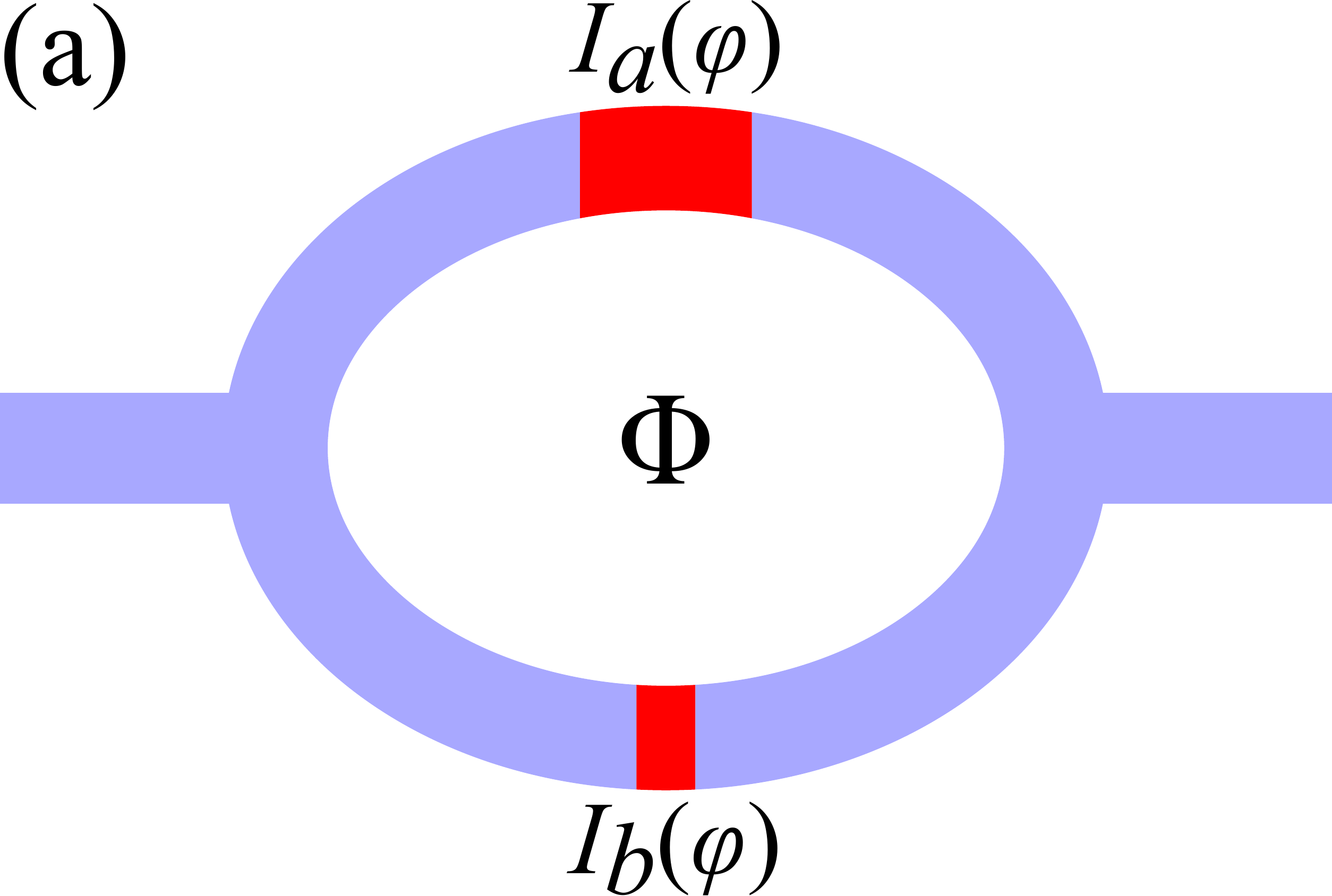}
 \\
 \vspace{3mm}
 \includegraphics[width=0.48\columnwidth]{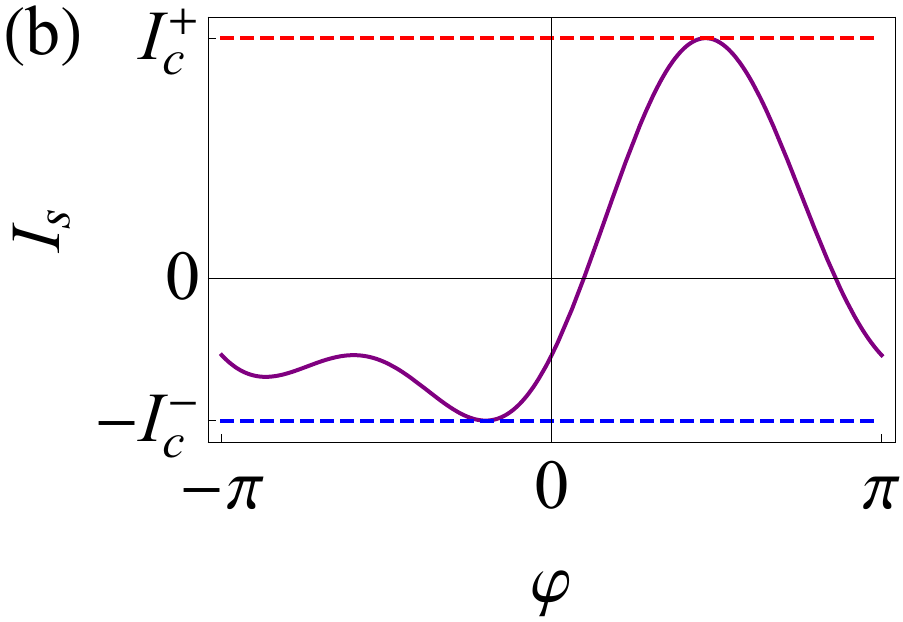}
 \hfill
 \includegraphics[width=0.46\columnwidth]{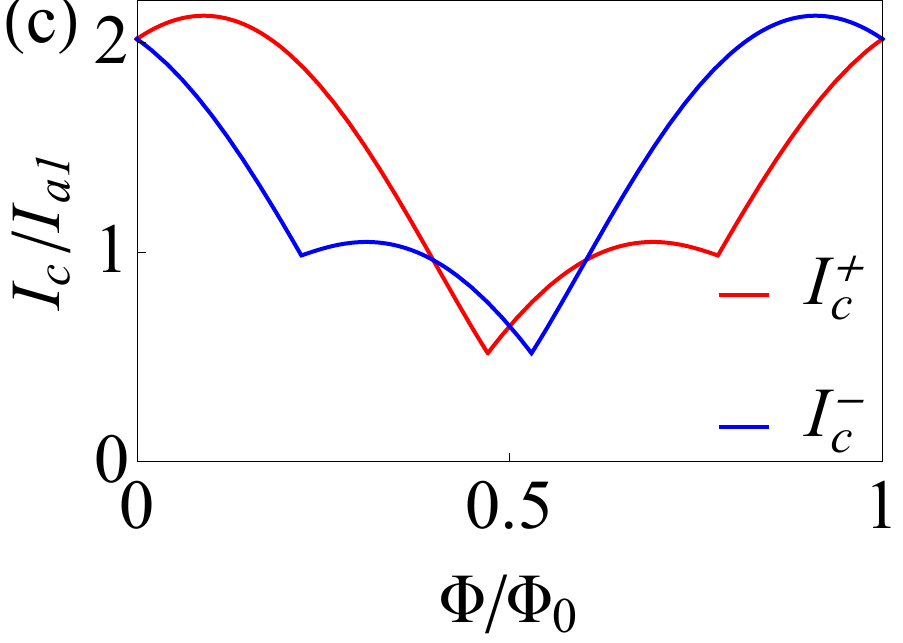}
\caption{(a)~Asymmetric SQUID with different  (generally nonsinusoidal) CPRs $I_a(\varphi)$ and $I_b(\varphi)$ in the two JJs. The ``minimal model'' demonstrating the JDE is defined by Eqs.\ \eqref{eq:Is_SQUID} and \eqref{eq:minmod}.
(b)~CPR for the ``minimal model'' of an asymmetric higher-harmonic SQUID, see Eq.\ \eqref{eq:J}, at $A=0.5$, $\tilde\phi=0.4\pi$. Note that $I_c^+ \neq I_c^-$, which is a manifestation of the JDE.
(c)~Magnetic interference pattern in the case of  $I_{b1}/I_{a1}=0.8$, $I_{b2}/I_{a1}=0.5$. The JDE is absent only at such values of the flux $\Phi$ that $\sin\tilde\phi=0$ (see Sec.~\ref{sec:discussion} for discussion of relation between the phase shift $\tilde\phi$ and the normalized flux $\phi$); those values correspond to crossings between the $I_c^+$ and $I_c^-$ curves. At all other values of $\Phi$, we find $I_c^+ \neq I_c^-$, which is a manifestation of the JDE. The magnetic field influences the diode efficiency (determined by $|I_c^+-I_c^-|$) and switches the polarity of the JDE (between positive one with $I_c^+>I_c^-$ and negative one with $I_c^+<I_c^-$).}
 \label{fig:SQUID}
\end{figure}

Additional versatility arises in SQUIDs, the Josephson systems of interferometer type \cite{BaroneBook,LikharevBook}. A system of this type is shown in Fig.~\ref{fig:SQUID}(a); the interferometer loop contains two JJs and is threaded by magnetic flux $\Phi$. The up-down asymmetry of such a system (asymmetry between junctions $a$ and $b$) in the presence of magnetic flux may lead to the left-right asymmetry for the current (the JDE). This is exemplified by SQUIDs with asymmetry of effective inductances included into the two interferometer arms \cite{Fulton1972.PhysRevB.6.855,Peterson1979,BaroneBook} or with an additional JJ included into one of the arms \cite{Gupta2022arXiv}.

Alternatively, the up-down asymmetry of the SQUID may be caused by higher Josephson harmonics in the (super)current-phase relation (CPR) of the JJs constituting the SQUID
\cite{MikhailovBSThesis2018,MikhailovMSThesis2020,Souto2022arXiv}.
In this paper, we develop analytical methods allowing us to describe the current-voltage characteristic (CVC) of asymmetric SQUIDs both in the absence and in the presence of external irradiation.
We analytically prove and characterize the JDE in asymmetric SQUIDs with higher Josephson harmonics, supporting our analytical results by numerical calculations.

The higher Josephson harmonics (contributions to the supercurrent of the form $\sin n\varphi$ with $n>1$, where $\varphi$ is the superconducting phase difference across a JJ) naturally arise in various types of JJs with not too low transparencies of their weak-link regions \cite{LikharevBook,Golubov2004review}.
Even if the two JJs in a SQUID are of the same type, they can have different CPRs due to variations of parameters if the weak link is represented by, e.g., a constriction (point contact) or a normal metal. Additional versatility is provided by JJs with ferromagnetic weak links, in which the first harmonic can be suppressed due to vicinity of the $0$-$\pi$ transition \cite{Ryazanov2001,Kontos2002,Sellier2003,Sellier2004,Oboznov2006}.
Recently, SQUIDs with high-transparency JJs have been realized and the presence of higher harmonics in their CPR has been experimentally demonstrated \cite{Nichele2020.PhysRevLett.124.226801}. This certifies that the system that we study can be realized in experiment. Importantly, realization of asymmetric higher-harmonic SQUIDs does not require exotic materials.
At the same time, alternatively, such SQUIDs can effectively be realized with the help of topological materials due  to asymmetry between the current-carrying edge states \cite{Chen2018.PhysRevB.98.075430}. In this case, the role of the two JJs in the SQUID is played by the two edge channels.

The paper is organized as follows:
In Sec.~\ref{sec:general}, we formulate general equations of the RSJ model suitable for describing the SQUIDs with higher Josephson harmonics.
In Sec.~\ref{sec:model}, we formulate the minimal model of an asymmetric higher-harmonic SQUID demonstrating the JDE. The model is characterized by sinusoidal contributions to the CPRs in both the JJs and, additionally, the second-harmonic contribution in one of them.
In Sec.~\ref{sec:CVClargej}, we analyze the CVC of our minimal model in the limit of large current.
In Sec.~\ref{sec:CVCsmallA}, we analyze the CVC in the limit of small second harmonic.
In Secs.~\ref{sec:spikes} and \ref{sec:steps}, we consider the Shapiro spikes and Shapiro steps, respectively, in the presence of external irradiation.
In Sec.~\ref{sec:discussion}, we discuss the obtained results.
In Sec.~\ref{sec:conclusions}, we present our conclusions.
Finally, some details of calculations are presented in the Appendices.

\section{General equations}
\label{sec:general}

\subsection{RSJ model, single sinusoidal junction}
\label{sec:RSJ}

In order to establish the reference point for our further calculations and to set up notation, we start by reviewing the standard resistively-shunted junction (RSJ) model \cite{BaroneBook,LikharevBook,TinkhamBook}. In its simplest form, this model represents a realistic Josephson junction (JJ) as the ideal JJ with the critical current $I_1$ shunted by the resistor with resistance $R$, and the CPR of the ideal JJ is assumed to be sinusoidal, $I_s(\varphi)=I_1 \sin\varphi$. The time dependence is assumed to enter only through $\varphi(t)$, which implies adiabatic approximation \cite{Larkin1966}.

The natural unit of frequency in this model is
\begin{equation} \label{eq:omega0}
\omega_0 = 2e I_1 R/\hbar,
\end{equation}
and we define dimensionless time, total current, and voltage bias as
\begin{equation} \label{eq:taujv}
\tau = \omega_0 t, \qquad j = I/I_1, \qquad v = V/I_1 R,
\end{equation}
respectively.

The second Josephson relation (between voltage and phase) and the RSJ equation (for the phase) then take the form
\begin{gather}
v = \dot \varphi,
\label{Vphi}
\\
\dot\varphi + \sin\varphi = j
\label{varphi_eq}
\end{gather}
(dot here stands for $d/d\tau$).
In the regime of fixed current, at $|j|<1$, we have the stationary Josephson effect, with time-independent phase determined from relation $\sin\varphi=j$.
At $|j|>1$, the Josephson effect is nonstationary (corresponding to the time-dependent $\varphi$), and Eq.\ \eqref{varphi_eq} can be integrated, which results in \cite{Aslamazov1969}
\begin{gather}
\varphi(\tau) = 2 \arctan \left[ \frac{1+ \nu \tan (\nu \tau/2)}{\sqrt{\nu^2+1}} \right] + 2\pi n,
\label{exactsol2}
\\
v(\tau) = \frac{j^2-1}{j+ \cos ( \nu \tau -\alpha )},
\label{eq:v(tau)}
\end{gather}
where
\begin{equation} \label{exactsol1}
\nu = \sqrt[\pm]{j^2-1},
\quad
\cos\alpha =  1/j,
\quad
\sin\alpha = \sqrt[\pm]{j^2-1}/j.
\end{equation}
Here, we have used our freedom to choose the integration constant, fixing the initial time (or, equivalently, fixing the phase of the Josephson oscillations).
In Eq.\ \eqref{exactsol1}, we have used notation
\begin{equation} \label{eq:sqrtpm}
\sqrt[\pm]{j^2-1} \equiv \sgn(j) \sqrt{j^2-1},
\end{equation}
where $\sqrt{\dots}$ is a positive quantity. This sign convention allows us to consider the CVC \cite{Stewart1968,McCumber1968,Aslamazov1969}
\begin{equation} \label{eq:overline_v}
\overline{v} = \nu
\end{equation}
[where the overline stands for time-averaging of $v(\tau)$] as an odd function defined both at positive and negative $j$.

Note that while Eq.\ \eqref{eq:overline_v} can be obtained directly as a result of time averaging of Eq.\ \eqref{eq:v(tau)}, this relation is actually a general relation for periodic processes.
Indeed, if we have a periodic variation of $v$ with period $\mathcal T$ and frequency $\nu=2\pi/\mathcal T$ (so that after time $\mathcal T$ the bias $v$ returns to its initial value, while $\varphi$ changes by $2\pi$), then from the general Josephson relation \eqref{Vphi} we find
\begin{equation} \label{eq:overline_v_general}
\overline{v} = \frac 1{\mathcal T} \int_0^{\mathcal T} v d\tau = \left. \frac 1{\mathcal T} \varphi(\tau) \right|_0^{\mathcal T} = \nu.
\end{equation}

In Eq.\ \eqref{exactsol2}, the arctangent is $2\pi/\nu$-periodic while the overall growth of $\varphi(\tau)$ with time is described by the staircase function
\begin{equation} \label{eq:n(tau)}
n(\tau)= \left\lfloor \nu\tau/2\pi \right\rceil,
\end{equation}
where $\left\lfloor \ldots \right\rceil$ denotes rounding to the nearest integer.

\subsection{Higher harmonics}

In many physical situations, due to the presence of high-transmission conducting channels in the weak link, the CPR of a JJ can deviate from sinusoidal due to higher Josephson harmonics \cite{LikharevBook,Golubov2004review}:
\begin{equation}
I_s(\varphi) = \sum_{n=1}^{\infty} I_n \sin n\varphi.
\end{equation}
This is actually so even in the case of the standard tunneling junction if higher orders with respect to the interface transparency are taken into account \cite{Kupriyanov1992,Golubov2005,Osin2021.PhysRevB.104.064514}.

In the presence of higher Josephson harmonics, if we still normalize the current to the amplitude of the first harmonic, Eq.\ \eqref{varphi_eq} should be substituted by
\begin{equation} \label{varphi_eq_J}
\dot\varphi + J(\varphi) = j,
\qquad
J(\varphi) = I_s(\varphi) / I_1.
\end{equation}
At the same time, Eqs.\ \eqref{eq:omega0}--\eqref{Vphi}, \eqref{eq:overline_v}, and \eqref{eq:overline_v_general} are left intact.

\subsection{SQUID}

A two-junction dc SQUID is a parallel connection of two JJs (indexed below as $a$ and $b$) \cite{TinkhamBook}, see Fig.~\ref{fig:SQUID}(a). This type of interferometer is sensitive to the magnetic flux $\Phi$ threading the superconducting loop.
The flux leads to the difference between the phase jumps at the two JJs:
\begin{equation} \label{eq:phi}
  \varphi_a-\varphi_b = \phi,\qquad \phi = 2\pi \Phi/\Phi_0.
\end{equation}
Defining $\varphi$ as the average of the two phase jumps, we can write the CPR of the SQUID as
\begin{equation} \label{eq:Is_SQUID}
   I_s(\varphi,\phi) = I_a(\varphi_a) + I_b(\varphi_b)  = I_a(\varphi+\phi/2) + I_b(\varphi-\phi/2).
\end{equation}
The CPR of each individual JJ is generally a sum of the Josephson harmonics,
\begin{gather}
  I_{a(b)}(\varphi) = \sum_{n=1}^{\infty} I_{a(b)n} \sin n\varphi .
\end{gather}

In the simplest case of sinusoidal JJs, only the first harmonics are retained. Then
\begin{align}
I_s(\varphi,\phi) &= I_{a1} \sin\left(\varphi+ \phi/2 \right) + I_{b1} \sin\left(\varphi-\phi/2 \right)
\notag \\
&= I_1 (\phi) \sin(\varphi+\gamma),
 \label{eq:I_SQUID}
\end{align}
where the critical current of the SQUID (the amplitude of the first harmonic) and the phase shift are determined by the following relations:
\begin{gather}
I_1 (\phi) = \sqrt{I_{a1}^2 + I_{b1}^2 +2 I_{a1} I_{b1} \cos\phi},
\label{eq:I_1(phi)}
\\
\tan \gamma = \frac{I_{a1}-I_{b1}}{I_{a1}+I_{b1}} \tan\frac{\phi}{2}.
\label{eq:beta}
\end{gather}

The effective CPR \eqref{eq:I_SQUID} is still sinusoidal (while the phase can be shifted in order to eliminate $\gamma$).
Therefore, the RSJ formulas of Sec.~\ref{sec:RSJ} are valid with the only remark that the first-harmonic amplitude is now flux-dependent according to Eq.\ \eqref{eq:I_1(phi)}.

\section{Model: asymmetric higher-harmonic SQUID as a Josephson diode}
\label{sec:model}

In any JJ, one can define the critical currents $I_c^+$ and $I_c^-$ corresponding to the two opposite directions of the current flow.
In a single JJ, $I_c^+ = I_c^-$. At the same time, in a SQUID, at $\phi\neq 0 \bmod{\pi}$, the two critical currents are different in the general case. A system with $I_c^+ \neq I_c^-$ can be called a superconducting Josephson diode as this implies nonreciprocal properties of the supercurrent.

Actually, in a SQUID, $I_c^+ = I_c^-$ only in certain special cases, e.g.,
(a) in a symmetric SQUID with $I_a(\varphi)=I_b(\varphi)$ (at arbitrary number and amplitudes of the harmonics),
(b) in the case when $I_a(\varphi)$ and $I_b(\varphi)$ are both described by the same single harmonic (with arbitrary amplitudes in the two junctions).

The most practical example of case (b) is when $I_a(\varphi)$ and $I_b(\varphi)$ are both sinusoidal. The SQUID can then be asymmetric, so that the critical currents of the two JJ are different; the SQUID is anyway characterized by $I_c^+=I_c^-$ in this case. In order to consider nonreciprocal effect with direction-dependent critical current, we consider the simplest asymmetric higher-harmonic SQUID with
\begin{equation} \label{eq:minmod}
I_a(\varphi) = I_{a1} \sin\varphi,
\quad
I_b(\varphi) = I_{b1} \sin\varphi + I_{b2} \sin 2\varphi.
\end{equation}
Different harmonic content of the two CPRs is already sufficient for the JDE. So, inclusion of the higher (second) harmonic in addition to the first one in only one of the junctions can be considered as a ``minimal model'' of the SQUID diode \cite{MikhailovBSThesis2018,MikhailovMSThesis2020,Souto2022arXiv}, see Fig.~\ref{fig:SQUID}.

We normalize the CPR of the SQUID, Eq.\ \eqref{eq:Is_SQUID}, by the amplitude of the first harmonic, Eq.\ \eqref{eq:I_1(phi)}.
We also shift the phase $\varphi$ by $\gamma$, so that the sum of the first harmonics produces $\sin\varphi$ after normalization, see Eq.\ \eqref{eq:I_SQUID} (the shift can be considered as a redefinition of the phase, which is anyway not important in a dynamical problem in the presence of voltage bias).
As a result, we obtain the RSJ equation in the form of Eq.\ \eqref{varphi_eq_J} with
\begin{equation} \label{eq:J}
J(\varphi) = \sin \varphi + A \sin(2\varphi - \tilde\phi),
\end{equation}
where
\begin{equation} \label{eq:Atildephi}
A(\phi) = I_{b2} / I_1(\phi),
\quad
\tilde\phi = \phi + 2\gamma(\phi).
\end{equation}
Here, $\tilde\phi$ plays the role of effective phase shift between the two Josephson harmonics.
The diode effect is present if
$\tilde\phi \neq 0 \bmod{\pi}$,
i.e., if $\tilde\phi$ is not an integer multiple of $\pi$, see Fig.~\ref{fig:SQUID}(c).
In particular, this implies $\phi\neq 0 \bmod{\pi}$.
In the case of equal first harmonics, $I_{a1}=I_{b1}$, the effective phase shift coincides with the normalized flux, $\tilde\phi=\phi$.

Our model thus requires solving Eqs.\ \eqref{Vphi}, \eqref{eq:overline_v}, \eqref{varphi_eq_J}, and \eqref{eq:J}.
Various quantities entering those equations are, in turn, defined in Eqs.\ \eqref{eq:omega0}, \eqref{eq:taujv}, \eqref{eq:phi}, \eqref{eq:I_1(phi)}, \eqref{eq:beta}, and \eqref{eq:Atildephi}.

Interestingly, the same form of the CPR, Eq.\ \eqref{eq:J}, effectively arises in different physical systems such as combined $0$-$\pi$ JJs in magnetic field \cite{Goldobin2011.PhysRevLett.107.227001}, JJs between exotic superconductors with broken time-reversal symmetry \cite{Trimble2021}, and SQUIDs with three sinusoidal JJs in the loop \cite{Gupta2022arXiv}. The asymmetry between $I_c^+$ and $I_c^-$, and also asymmetry of the magnetic diffraction pattern have been discussed in those contexts.

\begin{figure}[t]
 \includegraphics[width=\columnwidth]{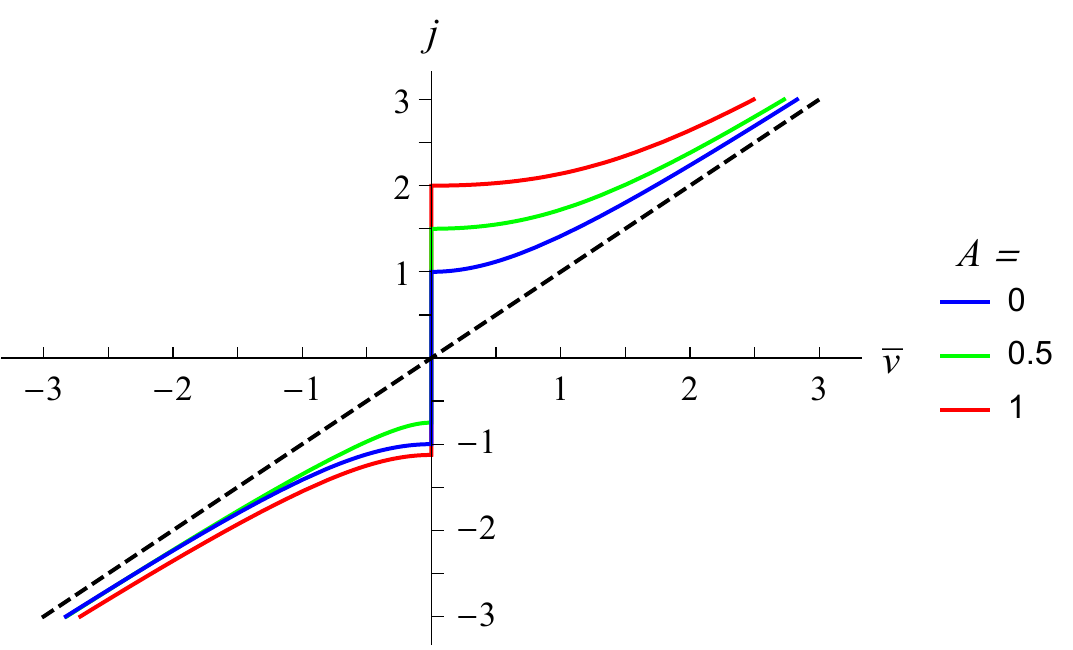}
\caption{CVC of the ``minimal model'', see Eq.\ \eqref{eq:J}, of the asymmetric higher-harmonic SQUID at $\tilde\phi=\pi/2$ and at different amplitudes $A$ of the second harmonic. The dashed line corresponds to the ohmic law $j=\overline{v}$. The asymmetry of the CVC curves, $j(-\overline{v}) \neq -j(\overline{v})$, is a manifestation of the JDE.}
 \label{fig:CVC}
\end{figure}

The CVC of our minimal model can be calculated numerically, see Fig.~\ref{fig:CVC}. In the next two sections, we develop analytical theory that provides results in the limits cases of either large $j$ or small $A$.

\section{CVC: harmonic perturbation theory in the limit of large current, \texorpdfstring{$|j|\gg 1$}{|j|>>1}}
\label{sec:CVClargej}

When the CPR is nonsinusoidal, the RSJ equations cannot be explicitly integrated. In this case, we can search for approximate solutions in certain limiting cases. One important limiting case is when the current is large, $|j|\gg 1$.

In order to formulate the method capable of treating this limiting case, we first consider the sinusoidal-junction case, and then switch to the general case.
Of course, in the sinusoidal case, the exact solution is known at any $j$ [see Eqs.\ \eqref{exactsol2}--\eqref{eq:sqrtpm}], so we consider this case simply as a testing ground for our method.

Note that our model of asymmetric higher-harmonic SQUID, see Eq.\ \eqref{eq:J}, reduces to the sinusoidal case of Eq.\ \eqref{varphi_eq} at $A=0$.
Straightforward perturbation theories often do not work for this type of equations. The problem is that besides oscillations, the solution contains a contribution linearly growing with time; this contribution corresponds to the average voltage bias, see Eq.\ \eqref{Vphi}. Any inaccuracy in finding the slope of this linear part immediately leads to linear growth of corrections (calculated in the framework of a straightforward perturbation theory and assumed to be small).
We therefore have to develop a perturbation theory which is applicable to Eq.\ \eqref{varphi_eq} at $|j|\gg 1$.
Our approach is a development of the idea proposed by Volkov and Nad' \cite{Volkov1970} (a similar idea has also been used in the context of the sine-Gordon equation \cite{Salerno1999.PhysRevB.59.14653}).

\subsection{Sinusoidal junction}
\label{sec:sinharm}

The general representation for the phase in the case of periodic $v(\tau)$ and/or $j(\tau)$ with period $2\pi/\nu$ is
\begin{equation} \label{phigen}
\varphi(\tau) = \nu \tau + \frac{a_0}2
+ \sum_{n=1}^\infty \left( a_n \cos n\nu\tau + b_n \sin n\nu\tau \right).
\end{equation}
This form automatically provides correspondence between the slope of the linear part of $\varphi(\tau)$ and frequency of its periodic part.
For convenience, we shift the time origin so that $a_0=0$.

Here, we develop the perturbation theory that sequentially takes into account higher harmonics, i.e., terms with $n>1$ in Eq.\ \eqref{phigen}. This approach works if $a_n$ and $b_n$ decrease with increasing $n$. This is so, e.g., in the limit of large fixed current, $|j|\gg 1$.
In this limit, $\nu$ and $a_n$, $b_n$ at $n\geqslant 1$, are represented as series with respect to $j^{-1}$.
We call this approach the ``harmonic'' perturbation theory.

In order to test our approach, we consider four orders of the perturbation theory (see Appendix~\ref{sec:harmsindetails} for detail). The result for the frequency immediately implies the result for the CVC:
\begin{equation} \label{eq:nuharm}
\overline{v} = \nu = j- 1/2j - 1/8j^3 .
\end{equation}
This reproduces the expansion of the exact result $\nu = \sqrt[\pm]{j^2-1}$, as it should.

\subsection{Asymmetric higher-harmonic SQUID}

The harmonic perturbation theory formulated above can also be applied in the general case when the CPR $\sin\varphi$ is substituted by arbitrary phase dependence $J(\varphi)$.
In the particular case of the asymmetric higher-harmonic SQUID defined by Eq.\ \eqref{eq:J}, the phase equation \eqref{varphi_eq_J}
takes the form
\begin{equation} \label{eqstart1b}
\dot\varphi + \sin\varphi + A \sin (2\varphi-\tilde\phi) = j.
\end{equation}
Considering three orders of the perturbation theory (see Appendix~\ref{sec:harmnonsindetails} for detail), we find
\begin{equation} \label{eq:nulargej}
\overline{v} = \nu =j- (1+A^2)/2j - 3A \sin(\tilde\phi) / 4j^2.
\end{equation}

The last term in Eq.\ \eqref{eq:nulargej} yields asymmetry of the CVC: due to nonzero $A$, the CVC $\overline{v}(j)$ is not an odd function any more, and the even contribution depends on the effective phase shift (hence, it is flux-dependent). This feature is a manifestation of the JDE at $|j|\gg 1$ (note that the amplitude $A$ of the second Josephson harmonic can be arbitrary).

\section{CVC: perturbation theory with respect to small second Josephson harmonic (\texorpdfstring{$|A|\ll 1$}{|A|<<1})}
\label{sec:CVCsmallA}

Now we concentrate on finding the correction to the CVC \eqref{eq:overline_v} due to small second harmonic ($|A|\ll 1$) at arbitrary current $j$.
In this case, we can employ a modification of the method proposed by  Thompson \cite{Thompson1973} in the framework of the Shapiro steps problem.
We can treat the second-harmonic perturbation (in the absence of external irradiation) in the same manner as the perturbation due to the external irradiation, applying the general idea of perturbation theory with feedback \cite{Thompson1973}. Technically, the feedback solves the problem of corrections which grow with time.

In order to illustrate the JDE, below we consider the first-order perturbation with respect to $A$.
An alternative technique capable of achieving the same goal is described in Appendix~\ref{sec:app:direct}. This technique can also be useful for treating higher orders of the perturbation theory.

\subsection{Perturbation theory with feedback}
\label{sec:alaThompsonPhi}

In order to solve Eq.\ \eqref{eqstart1b},
we represent $\varphi(\tau)$ and $j$ as series with respect to small $A$:
\begin{align}
\varphi(\tau) &= f_0(\tau) + A f_1(\tau) + A^2 f_2(\tau) + \dots,
\\
j &= i_0 + A i_1 + A^2 i_2 + \dots
\end{align}
Order by order, Eq.\ \eqref{eqstart1b} yields
\begin{align}
& \dot f_0 + \sin f_0 = i_0,
\label{eq:varphi0} \\
& \dot f_n + f_n \cos f_0 = D_n,
\quad
n=1,2,\dots,
\label{phinb}
\end{align}
with the driving function
$D_n = i_n + F_n$,
where
\begin{align}
F_1 &= -\sin (2 f_0-\tilde\phi),
\\
F_2 &= (f_1^2/2) \sin f_0 -2 f_1\cos (2 f_0-\tilde\phi), \quad \dots
\end{align}
It is important that $F_n$ and hence $D_n$ contain only those $f_k$ that have $k<n$, which are already known from previous orders of the perturbation theory.

The zeroth-order equation, Eq.\ \eqref{eq:varphi0}, can obviously be obtained from Eq.\ \eqref{varphi_eq} as a result of substitution $\varphi\mapsto f_0$, $j\mapsto i_0$. Its solution is given  by Eqs.\ \eqref{exactsol2} and \eqref{eq:n(tau)} with the corresponding modifications to $\nu$ and $\alpha$, which now acquire the $0$ subscript and take the form $\nu_0 =\sqrt[\pm]{i_0^2-1}$ and $\cos\alpha_0 = 1/i_0$ instead of Eq.\ \eqref{exactsol1}.

Our expansion assumes that $f_n(\tau)$ with $n=1,2,\dots$ are limited functions.
Since $f_n(\tau)$ with $n=1,2,\dots$ do not grow with time, time averaging of Eq.\ \eqref{Vphi} yields
\begin{equation}
\overline{v} =  \overline{\dot f_0} =  \sqrt[\pm]{i_0^2-1}.
\end{equation}
We are interested in finding $j$ vs $\overline{v}$, which becomes $(i_0 + A i_1 + A^2 i_2 + \dots)$ vs $\sqrt[\pm]{i_0^2-1}$.

Since Eq.\ \eqref{phinb} is a first-order linear differential equation, it can be integrated as follows:
\begin{align}
f_n(\tau) &= \frac 1{K(\tau)} \int_{\tau_n}^\tau D_n(t) K(t) dt,
\\
K(\tau) &= \exp \int_0^\tau \cos f_0(t) dt,
\end{align}
where $\tau_n$ are arbitrary integration constants.
The last integration can be performed with the help of relation $\cos f_0(\tau) = -  \ddot f_0 / \dot f_0$ which follows from Eq.\ \eqref{eq:varphi0}. The result is
\begin{equation}
K(\tau) = \frac{1+\cos\alpha_0 \cos\theta(\tau)}{1+\cos^2 \alpha_0},
\quad
\theta(\tau) = \nu_0\tau-\alpha_0.
\end{equation}
The consistency of our perturbative procedure requires that $f_n(\tau)$ does not grow with time, which implies $\overline{D_n(\tau) K(\tau)} = 0$.
In the case of $n=1$, this yields
\begin{equation}
\overline{\bigl[ i_1 - \sin \bigl( 2 f_0(\tau) -\tilde\phi \bigr) \bigr] \bigl[ 1+\cos\alpha_0 \cos\theta(\tau) \bigr]} = 0,
\end{equation}
which results in
\begin{equation}
  i_1  = \Bigl[ 1-2 (i_0^2-1) \Bigl( 1- \sqrt[\pm]{i_0^2-1} /i_0 \Bigr) \Bigr] \sin\tilde\phi.
\end{equation}

As a result, in the first order with respect to $A$, we have parametric dependence $\overline{v}(i_0) = \sqrt[\pm]{i_0^2-1}$, $j(i_0) = i_0 + A i_1(i_0)$.
In this order, the physically measurable direct $\overline{v}(j)$ dependence takes the form
\begin{equation} \label{eq:vA}
  \overline{v} =\sqrt[\pm]{j^2-1} - A \frac{2 \sqrt[\pm]{j^2-1}^3 -2j^3 +3 j }{\sqrt[\pm]{j^2-1}} \sin\tilde\phi.
\end{equation}
In the limit $|j|\gg 1$, this result reproduces the leading $A$ term in Eq.\ \eqref{eq:nulargej}.

At $\tilde\phi \neq 0\bmod{\pi}$, the CVC given by Eq.\ \eqref{eq:vA} is asymmetric. While the first (main) term is an odd function of $j$, the second term (the $A$ correction) is even.
This is a manifestation of the JDE.

Later, we will also need the explicit result for $f_1(\tau)$, which can be written as
\begin{equation} \label{eq:varphi1Thompson}
  f_1(\tau) = f_{1s}(\tau) + f_{1c}(\tau) + f_{\tau_1}(\tau),
\end{equation}
where
\begin{gather}
  f_{1s}(\tau) = \frac{2 g_s (\tau) \sin\tilde\phi}{1+\cos\alpha_0 \cos\theta(\tau)},
  \;
  f_{1c}(\tau) = \frac{2 g_c (\tau) \cos\tilde\phi}{1+\cos\alpha_0 \cos\theta(\tau)},
  \\
  f_{\tau_1}(\tau) = -2\frac{g_s(\tau_1) \sin\tilde\phi + g_c(\tau_1) \cos\tilde\phi}{1+\cos\alpha_0 \cos\theta(\tau)},
\end{gather}
and
\begin{align}
&g_s(\tau) =
- (2-\sin\alpha_0) \sin\alpha_0 \sin\theta(\tau)
\notag \\
&+\frac{\sin^2 \alpha_0}{\cos \alpha_0} \left[
\theta(\tau) - 2\pi n_{\theta} -2 \arctan\left( \tan\frac{\alpha_0}{2} \tan\frac{\theta(\tau)}{2} \right)
\right],
\\
&g_c(\tau) = \sin \alpha_0 \tan\alpha_0 \ln \left[ 1+\cos\alpha_0 \cos\theta(\tau) \right] - \cos\theta(\tau),
\label{eq:g_c}
\end{align}
with $n_\theta = \left\lfloor \theta(\tau)/2\pi \right\rceil$.
Both $g_s$ and $g_c$ functions are $2\pi$-periodic with respect to $\theta$.

\section{Shapiro spikes}
\label{sec:spikes}

In the presence of external irradiation of frequency $\omega$, the JJ (rf-driven JJ) demonstrates peculiarities of the CVC at certain voltages at which the internal Josephson oscillations are synchronized with external irradiation (so-called phase locking).
The simplest example of this kind is represented by the \emph{voltage-driven} junction: $V(t) = V_\mathrm{dc} + V_\mathrm{ac} \cos\omega t$ \cite{BaroneBook,Gregers-Hansen1972}. In dimensionless units of Eq.\ \eqref{eq:taujv}, we have
\begin{equation}
v(\tau) = v_\mathrm{dc} + v_\mathrm{ac} \cos\Omega \tau,
\end{equation}
where $\Omega = \omega/\omega_0$.
Direct integration of Eq.\ \eqref{Vphi} then yields
\begin{equation}
\varphi(\tau) = \varphi_0 + v_\mathrm{dc} \tau + (v_\mathrm{ac}/\Omega) \sin \Omega \tau.
\end{equation}

In the most general case of the normalized CPR of the form
\begin{equation}
J(\varphi) = \sum_{k=1}^{\infty} A_k \sin (k\varphi + \Phi_k),
\end{equation}
the time dependence of the supercurrent can then be written as
\begin{multline} \label{eq:Jkn}
J(\tau) = \sum_{k=1}^\infty \sum_{n=-\infty}^\infty (-1)^n A_k J_n \left( \frac{k v_\mathrm{ac}}\Omega \right)
\\
\times
\sin\left[ k\varphi_0 +\Phi_k + (k v_\mathrm{dc} - n\Omega) \tau \right]
\end{multline}
(see Appendix~\ref{sec:app:spikes} for detail).

The time-averaged total current of the RSJ model is then given by
\begin{multline}
\overline{j} =v_\mathrm{dc} + \overline{J}
= v_\mathrm{dc}
\\
+
\left\{
\begin{array}{ll}
{\sum\limits_{k,n}}' (-1)^n A_k J_n \left( \frac{k v_\mathrm{ac}}\Omega \right) \sin ( k\varphi_0 + \Phi_k), & \text{if } v_\mathrm{dc} = \frac{n}{k} \Omega,
\\
0, & \text{else}.
\end{array}
\right.
\end{multline}
The prime sign implies that the double sum is taken only over such pairs of $k$ and $n$ that correspond to relation $v_\mathrm{dc} = (n/k) \Omega$ (at given $v_\mathrm{dc}$).
The $\varphi_0$ phase can be arbitrary, so plotting the CVC $\overline{j}(v_\mathrm{dc})$, we obtain the linear Ohm's law with superimposed vertical spikes at those $v_\mathrm{dc}$ --- the Shapiro spikes.

In the simplest case of the sinusoidal CPR (with $A_k=\delta_{1k}$ and $\Phi_k=0$), only the integer Shapiro spikes (corresponding to $k=1$) are present. Their height is $2 J_{|n|} (v_\mathrm{ac} /\Omega)$, and the spikes are symmetric (their centers lie on the ohmic line $\overline{j} =v_\mathrm{dc}$).

In the general case, due to the $\Phi_k$ phase shifts, the spikes become asymmetric (their centers are shifted from the ohmic line). Moreover, asymmetry of the pairs of spikes corresponding to $\pm v_\mathrm{dc}$ appears, which implies that the overall $\overline{j}(v_\mathrm{dc})$ plot (with Shapiro spikes) is not an odd function, which is a manifestation of the JDE.

The above consideration is general and does not require smallness of the higher Josephson harmonics. In order to illustrate it in a simpler case, we consider the case of Eq.\ \eqref{eq:J} corresponding to two harmonics (the amplitude $A$ of the second harmonic can be arbitrary). Due to the presence of the second harmonic, the fractional half-integer Shapiro spikes (with $k=2$) appear. They are symmetric with respect to the ohmic line, and the maximum and minimum points for the spike at $v_\mathrm{dc} = (n/2) \Omega$ (with odd $n$) are given by
\begin{equation}
j_\mathrm{max(min)} = v_\mathrm{dc} \pm \left| A J_n (2v_\mathrm{ac}/\Omega) \right|.
\end{equation}
Note that this result does not depend on the effective phase shift $\tilde\phi$ and, hence, on the magnetic flux $\phi$.

At the same time, the integer Shapiro spikes at $v_\mathrm{dc} = n \Omega$ are contributed to by two pairs of $(k,n)$ corresponding to the two harmonics, $(1,n)$ and $(2,2n)$:
\begin{multline} \label{eq:intspike}
j_\mathrm{max(min)} = v_\mathrm{dc} + \underset{\varphi_0}{\maxmin} \Bigl[ J_{|n|} (v_\mathrm{ac}/\Omega) \sin\varphi_0
\\
+ A J_{|2n|} (2v_\mathrm{ac}/\Omega) \sin(2\varphi_0-\tilde\phi) \Bigr].
\end{multline}
At $\tilde\phi\neq 0 \bmod{\pi}$, the absolute values of the maximum and minimum operation in the right-hand side (rhs) of this expression do not coincide, which implies asymmetric vertical shift of the integer spikes (with respect to the ohmic line). At the same time, inversion of $v_\mathrm{dc}$ (corresponding to inversion of $n$) does not change the values of the maximum and minimum, so the vertical shift does not change (i.e., it is not inverted). This is a manifestation of the JDE since the overall $\overline{j}(v_\mathrm{dc})$ plot (with Shapiro spikes) is not an odd function.

The vertical shift takes place, in particular, in the case of the zeroth Shapiro spike corresponding to the supercurrent at $v_\mathrm{dc}=0$, implying that the absolute value of the critical current in the case of $\tilde\phi\neq 0 \bmod{\pi}$ is  direction-dependent.

\begin{figure}[t]
 \includegraphics[width=\columnwidth]{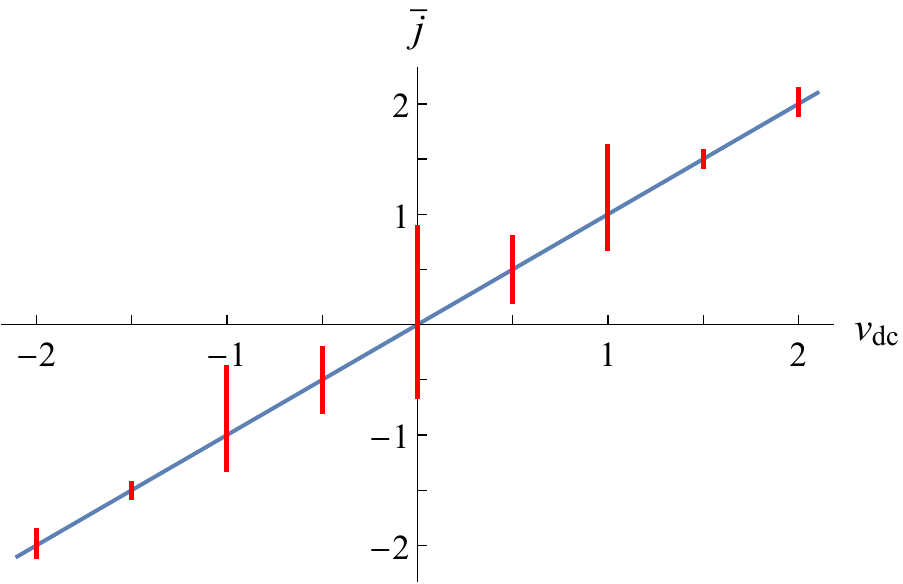}
\caption{CVC in the presence of external ac signal (voltage-driven case), demonstrating the Shapiro spikes (red vertical segments)  in the case of $A=0.5$, $\Omega=1$, $v_\mathrm{ac}=1$, $\tilde\phi=\pi/2$. The half-integer Shapiro spikes (at half-integer $v_\mathrm{dc}$), arising due to the second harmonic of the CPR, are symmetric. The integer Shapiro spikes (at integer $v_\mathrm{dc}$) are asymmetric due to nontrivial phase shift $\tilde\phi$ between the two harmonics: each integer spike is vertically shifted with respect to the ohmic line, and the shift is \emph{the same} for positive and negative voltage biases. This makes the whole picture asymmetric, $\overline{j}(-v_\mathrm{dc}) \neq -\overline{j}(v_\mathrm{dc})$, and illustrates the JDE.}
 \label{fig:spikes}
\end{figure}

The above results are illustrated by Fig.~\ref{fig:spikes}.

\section{Shapiro steps}
\label{sec:steps}

A different kind of CVC peculiarities, the Shapiro steps \cite{Shapiro1963.PhysRevLett.11.80,BaroneBook}, arise in the case of a more experimentally relevant \emph{current-driven} JJ with
$I(t) = I_\mathrm{dc} + I_\mathrm{ac} \cos(\omega t +\beta)$ (it turns out to be illustrative to keep the arbitrary phase $\beta$ of the external drive).
In dimensionless units of Eq.\ \eqref{eq:taujv},
we need to consider the following equation:
\begin{equation} \label{eqstart1gen}
\dot\varphi + J(\varphi) = j_\mathrm{dc} + j_\mathrm{ac} \cos (\Omega \tau +\beta).
\end{equation}

As pointed out in Ref.\ \cite{Volkov1970}, straightforward perturbation theory with respect to $j_\mathrm{ac}$ fails for this equation due to resonances (at $\Omega$ corresponding to the Shapiro steps).
In the limit $j_\mathrm{dc} \gg 1$, the authors of Ref.\ \cite{Volkov1970} proposed a nonperturbative method for finding the form of the CVC near the Shapiro steps. It was shown that the CVC near the Shapiro step had the same form as the CVC at $V = 0$ in the absence of irradiation.

A perturbative approach capable of treating the resonances was proposed by Thompson \cite{Thompson1973} in the case of the sinusoidal JJ with $J(\varphi) = \sin\varphi$.
At the same time, the method is actually rather general and can be applied to any CPR $J(\varphi)$. Below, we describe such generalization.

\subsection{Perturbation theory with feedback to treat resonances}
\label{sec:pertshap}

In Sec.~\ref{sec:alaThompsonPhi}, we have already applied the ideas of Thompson \cite{Thompson1973} to construct a divergence-free perturbation theory in the equation without external driving.
Now we apply the same ideas in the situation very close to the original formulation, i.e., in the presence of external driving, see Eq. \eqref{eqstart1gen}.
The difference from the work by Thompson  \cite{Thompson1973} is that we assume arbitrary CPR $J(\varphi)$ instead of the sinusoidal one.

We represent $\varphi(\tau)$ and $j_\mathrm{dc}$ as series with respect to small $j_\mathrm{ac}$:
\begin{align}
\varphi(\tau) &= \varphi_0(\tau) + \varphi_1(\tau) + \varphi_2(\tau) + \dots,
\\
j_\mathrm{dc} &= j_0 + j_1 + j_2 + \dots
\end{align}
Order by order, Eq.\ \eqref{eqstart1gen} yields
\begin{align}
& \dot \varphi_0 + J(\varphi_0) = j_0,
\label{phi0gen}
\\
& \dot \varphi_n + \left[ \partial J(\varphi_0) / \partial \varphi_0 \right] \varphi_n = D_n,
\quad
n=1,2,\dots,
\label{phingen}
\end{align}
with the driving function
$D_n = j_n + F_n$,
where
\begin{align}
F_1 &= j_\mathrm{ac} \cos (\Omega \tau+\beta),
\\
F_2 &=  - \left[ \partial^2 J(\varphi_0) / \partial \varphi_0^2 \right]  \varphi_1^2 /2, \quad \dots
\end{align}
It is important that $F_n$ and hence $D_n$ contain only those $\varphi_k$ that have $k<n$, which are already known from previous orders of the perturbation theory.

Our expansion assumes that $\varphi_n(\tau)$ with $n=1,2,\dots$ are small.
Since $\varphi_n$ with $n=1,2,\dots$ do not grow with time,  time averaging of Eq.\ \eqref{Vphi} yields
\begin{equation} \label{eq:overlinev}
\overline{v} =  \overline{\dot \varphi_0}
\end{equation}
--- the latter quantity must be found from the solution of Eq.\ \eqref{phi0gen}.
We are interested in finding $j_\mathrm{dc}$ vs $\overline{v}$, which becomes $(j_0 + j_1 + j_2 + \dots)$ vs $\overline{\dot \varphi_0}$.

Since Eq.\ \eqref{phingen} is a first-order linear differential equation, it can be integrated as follows:
\begin{align}
\varphi_n(\tau) &= \frac 1{K(\tau)} \int_{\tau_n}^\tau D_n(t) K(t) dt,
\label{eq:varphin} \\
K(\tau) &= \exp \int_0^\tau \frac{\partial J(\varphi_0)}{\partial \varphi_0}(t) dt,
\end{align}
where $\tau_n$ are arbitrary constants.
Differentiating Eq.\ \eqref{phi0gen} with respect to $\tau$, we find
\begin{equation}
\frac{\partial J(\varphi_0)}{\partial \varphi_0}(\tau) = -\frac{\ddot \varphi_0(\tau)}{\dot \varphi_0(\tau)},
\end{equation}
hence
\begin{equation}
K(\tau) = \dot \varphi_0(0) / \dot \varphi_0(\tau).
\end{equation}

The consistency of our perturbative procedure requires that $\varphi_n(\tau)$ does not grow with time, which implies
\begin{equation} \label{eq:Dnphi0}
\overline{D_n(\tau) / \dot \varphi_0(\tau)} = 0.
\end{equation}
The algorithm is then as follows: We need to find the zeroth-order solution from Eq.\ \eqref{phi0gen}. After that, we take $n=1$, find $j_1$ from Eq.\ \eqref{eq:Dnphi0} and then find $\varphi_1(\tau)$ from Eq.\ \eqref{eq:varphin}. Then we do the same in the case of $n=2$, etc.

\subsection{The first Shapiro steps}
\label{sec:firstShapiro}

In Sec.~\ref{sec:pertshap}, we have developed the perturbation theory with respect to the ac external field ($j_\mathrm{ac}$).
In Eq.\ \eqref{eq:Dnphi0}, $\varphi_0(\tau)$ is the zeroth order of this perturbation theory (i.e, the solution in the absence of external field).
In the case of our asymmetric SQUID model, $\varphi_0(\tau)$ itself has been studied in Sec.~\ref{sec:alaThompsonPhi} by the perturbation theory with respect to small second harmonic ($A$).
Now we combine the two perturbation theories, writing $\varphi_0(\tau)$ in the first order with respect to $A$ as a sum of the sinusoidal-case solution and the correction of Eq.\ \eqref{eq:varphi1Thompson}:
\begin{equation} \label{eq:f0f1}
  \varphi_0(\tau) = f_0(\tau) + A f_1(\tau).
\end{equation}
In this order with respect to $A$, Eq.\ \eqref{eq:Dnphi0} in the case of $n=1$ takes the form
\begin{equation} \label{eq:average}
\overline{\frac{j_1 + j_\mathrm{ac} \cos (\Omega \tau+\beta)}{\dot f_0(\tau)} \biggl[ 1-A\frac{\dot f_1(\tau)}{\dot f_0(\tau)} \biggr]} = 0.
\end{equation}
The average voltage bias given by Eq.\ \eqref{eq:overlinev} is expressed in notation of Sec.~\ref{sec:alaThompsonPhi} as $\overline{v} = \nu_0 = \sqrt[\pm]{i_0^2-1}$.

Considering the sinusoidal-junction case ($A=0$) as a reference point, we obtain
\begin{equation} \label{eq:harmlock}
\overline{ [ j_1 + j_\mathrm{ac} \cos(\Omega \tau + \beta) ] [i_0 + \cos(\nu_0 \tau -\alpha_0)]} = 0.
\end{equation}
Except for the constant part $j_1 i_0$, nonzero contributions to the left-hand side (lhs) arise only
at $\nu_0 = \pm \Omega$ (phase locking of the oscillating cosines), which corresponds to the first Shapiro steps (positive and negative). In these cases, we obtain
\begin{equation} \label{eq:j1}
j_1 = -(j_\mathrm{ac} / 2i_0) \cos(\alpha_0 \pm \beta).
\end{equation}
The argument of the cosine is the phase difference between the internal Josephson oscillations and the external field. Since this relative phase can be arbitrary, the obtained result corresponds to the vertical segment of the $j_\mathrm{dc}(\overline{v})$ curve of total height $j_\mathrm{step} = j_\mathrm{ac}/\sqrt{\Omega^2+1}$. This first Shapiro step originates from the first order ($n=1$) of the perturbation theory of Sec.~\ref{sec:pertshap}.

Next, we take into account the $A$ correction (the second Josephson harmonic) in Eq.\ \eqref{eq:average}. The corresponding contribution originates from the $f_1$ function given by Eq.\ \eqref{eq:varphi1Thompson}. It turns out that for our discussion of the first Shapiro steps in the first order with respect to $A$, only the $f_{1s}$ term of $f_1$ is essential [see Eqs.\ \eqref{eq:varphi1Thompson}--\eqref{eq:g_c}].
It produces two contributions to Eq.\ \eqref{eq:j1} with the order of magnitude $A j_1$ and $A j_\mathrm{ac} \cos(\alpha_0 \pm \beta)$, while the $f_{1c}$ and $f_{\tau_1}$ terms produce contributions of the order of $A j_\mathrm{ac} \sin(\alpha_0 \pm \beta)$. Determining the height of the Shapiro step, we take $\cos(\alpha_0 \pm \beta) = \pm 1$ in the main order with respect to $A$, see Eq.\ \eqref{eq:j1}. The $A j_\mathrm{ac} \sin(\alpha_0 \pm \beta)$ corrections in the rhs of Eq.\ \eqref{eq:j1} slightly shift the extremal points [such that $\sin(\alpha_0 \pm \beta) \sim A$] but this only gives a $A^2$ correction to the height of the Shapiro step. Since we work in the first order with respect to $A$, we disregard this correction. So, the first-order correction is only produced by the $f_{1s}$ term of $f_1$.

Performing the corresponding calculation, we obtain Eq.\ \eqref{eq:j1} with corrections of the order of $A j_1$ and $A j_\mathrm{ac} \cos(\alpha_0 \pm \beta)$. Determining $j_1$ in the first order with respect to $A$ from the resulting equation, we find
\begin{multline}
  j_1 = - (j_\mathrm{ac} /2 i_0) \cos(\alpha_0 \pm \beta)
  \\
  \times  \Bigl\{
  1
  + A \sin\tilde\phi \Bigl[ (2i_0^2 -1 +2 i_0^{-2}) \sqrt[\pm]{i_0^2-1}-2i_0 (i_0^2-1) \Bigr]
  \Bigr\}.
\end{multline}
Finally, the height of the first Shapiro steps in the $j_\mathrm{dc}(\overline{v})$ dependence at $\overline{v} = \pm\Omega$ is
\begin{multline} \label{eq:Shapheight}
   j_\mathrm{step}^{(\pm 1)} = \frac{j_\mathrm{ac}}{\sqrt{\Omega^2+1}}
   \\
   \times \left(
  1 \pm A \Omega \frac{2\Omega^4 +3(\Omega^2+1) -2\Omega (\Omega^2+1)^{3/2}}{\Omega^2+1} \sin\tilde\phi
  \right).
\end{multline}

The heights of the two first Shapiro steps (positive and negative) are different, $j_\mathrm{step}^{(+1)} \neq j_\mathrm{step}^{(-1)}$. This is a manifestation of the JDE. The asymmetry appears due to the second harmonic of the CPR ($A\neq 0$) and nontrivial phase shift ($\sin\tilde\phi \neq 0$).
Due to the $\pm$ sign in Eq.\  \eqref{eq:Shapheight}, one of the steps is higher while the other one is lower than in the absence of the second Josephson harmonic.

\subsection{The \texorpdfstring{$1/2$}{1/2} Shapiro steps}

Due to the presence of the second harmonic in the CPR, fractional half-integer Shapiro steps  \cite{Gregers-Hansen1972,BaroneBook,Cuevas2002.PhysRevLett.88.157001,Chauvin2006.PhysRevLett.97.067006} should appear in the CVC
(note that fractional Shapiro steps can also arise in the case of purely sinusoidal CPR due to chaotic regimes \cite{Revin2018}; we do not consider such effects here).
The approach of Sec.~\ref{sec:firstShapiro} can be applied for calculating the first of them, the $1/2$ Shapiro steps at $\nu_0=\pm \Omega/2$.
This calculation takes into account the first orders with respect to $j_\mathrm{ac}$ and $A$, so Eqs.\ \eqref{eq:f0f1} and \eqref{eq:average} are still applicable.

The main Josephson harmonic does not produce phase locking in the case of fractional steps, so averaging in Eq.\ \eqref{eq:harmlock} now only yields the trivial contribution $j_1 i_0$.
Next, we take into account the $A$ correction (the second Josephson harmonic) in Eq.\ \eqref{eq:average}. The corresponding contribution originates from the $f_1$ function given by Eq.\ \eqref{eq:varphi1Thompson}.
In the case of subharmonic locking that we now discuss, the $f_{1s}$ and $f_{1c}$ terms of $f_1$ [see Eqs.\ \eqref{eq:varphi1Thompson}--\eqref{eq:g_c}] are essential while the contribution from $f_{\tau_1}$ vanishes as a result of averaging.
The $f_{1s}$ term produces two contributions to Eq.\ \eqref{eq:average} with the order of magnitude $A j_1$ and $A j_\mathrm{ac} \cos(2\alpha_0 \pm \beta)$, while the $f_{1c}$ term only produces a contribution of the order of $A j_\mathrm{ac} \sin(2\alpha_0 \pm \beta)$.
Determining $j_1$ in the first order with respect to $A$ from the resulting equation, we find
\begin{multline}
j_1  =
 \frac{2A j_\mathrm{ac}}3
 \left[
 (2i_0^2-3) \sqrt[\pm]{i_0^2-1} - \frac{2(i_0^2-1)^2}{i_0}
 \right]
 \\
 \times
\sin(\tilde\phi + 2\alpha_0 \pm \beta) .
\end{multline}
Finally, the height of the $1/2$ Shapiro steps in the $j_\mathrm{dc}(\overline{v})$ dependence at $\overline{v} = \pm\Omega/2$ is
\begin{equation}
   j_\mathrm{step}^{(\pm 1/2)} = \frac{2A j_\mathrm{ac}}3 \Omega \left[ 1-\frac{\Omega^2}{2} + \frac{\Omega^3}{2\sqrt{\Omega^2 + 4}} \right].
\end{equation}

In the main approximation we thus find steps of equal height, $j_\mathrm{step}^{(+1/2)} = j_\mathrm{step}^{(-1/2)}$, which does not depend on $\tilde\phi$ (this resembles the symmetry and flux-independence of the half-integer Shapiro \emph{spikes}, see Sec.~\ref{sec:spikes}).
However, since the underlying CVC (in the absence of the ac drive, i.e., at $j_\mathrm{ac}=0$) is asymmetric (see Sec.~\ref{sec:CVCsmallA}), it is natural to expect that the Shapiro steps are also generally asymmetric. This effect should take place in higher orders of the perturbation theory.

\subsection{Numerical results}

\begin{figure}[t]
 \includegraphics[width=\columnwidth]{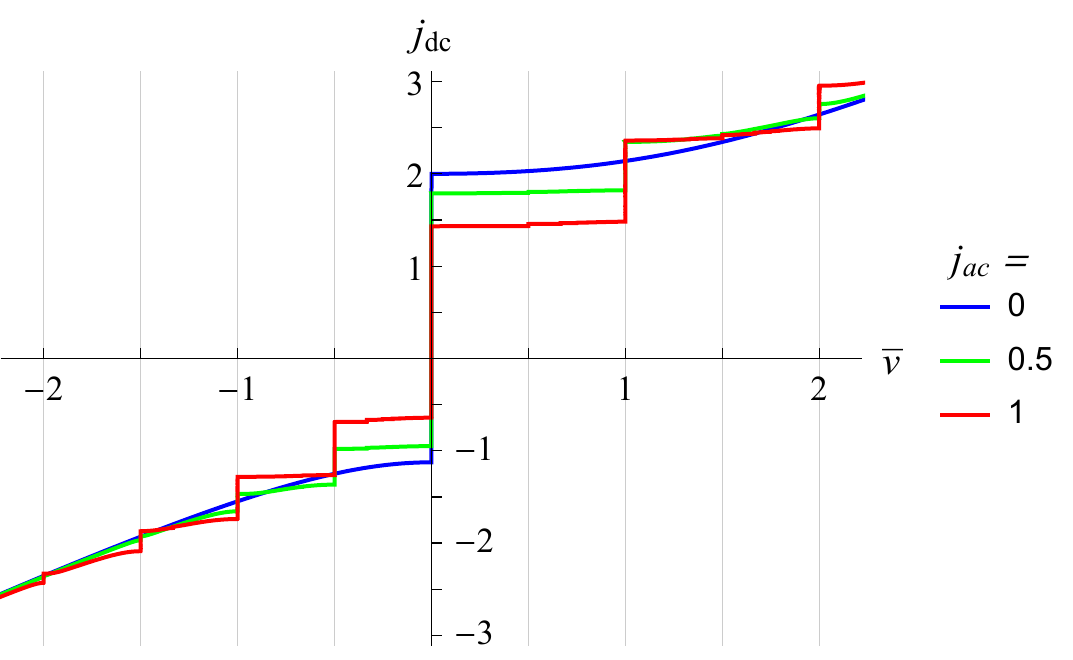}
\caption{CVC in the presence of external ac signal (current-driven case), demonstrating the Shapiro steps (vertical segments)  in the case of $A=1$, $\Omega=1$, $\tilde\phi=\pi/2$, and several values of $j_\mathrm{ac}$. Both integer and half-integer Shapiro steps (at integer and half-integer $\overline{v}$, respectively) are asymmetric due to the presence of the second harmonic of the CPR and nontrivial phase shift $\tilde\phi$ between the two harmonics. The asymmetry $j_\mathrm{dc}(-\overline{v}) \neq -j_\mathrm{dc}(\overline{v})$ illustrates the JDE.}
 \label{fig:steps}
\end{figure}

Figure~\ref{fig:steps} demonstrates asymmetry of both integer and half-integer Shapiro steps [taking place at $\overline{v}=(n/k)\Omega$ with integer and half-integer $(n/k)$] at values of $A$ beyond the perturbation theory applicability region.
In the case of the chosen parameters set, positive half-integer Shapiro steps at finite $j_\mathrm{ac}$ are very small, while the negative ones are clearly visible (and even larger than some of the neighboring integer steps).

Careful inspection of the obtained curves also evidences presence of fractional Shapiro steps with $(n/k)$ being rational fractions which are not half-integer, such as $1/3$, $2/3$, $1/4$, $3/4$, etc. However, they are usually smaller and less visible than the integer and half-integer steps.

\section{Discussion}
\label{sec:discussion}

As evidenced by our results given by Eqs.\ \eqref{eq:nulargej}, \eqref{eq:vA}, \eqref{eq:intspike}, and \eqref{eq:Shapheight}, the strength of the diode effect (and its polarity) is governed by the amplitude of the second Josephson harmonic multiplied by $\sin\tilde\phi$. In its turn, the effective phase shift $\tilde\phi$ between the Josephson harmonics is a function of the experimentally controllable (normalized) SQUID flux $\phi$. At the same time, the $\tilde\phi(\phi)$ dependence is nontrivial, see Eq.\ \eqref{eq:Atildephi}. This dependence qualitatively depends on the ratio between the first Josephson harmonics of the two junctions in the SQUID, $I_{b1}/I_{a1}$ [note that in our minimal model, $b$ labels the junction with the second Josephson harmonic, see Eq.\ \eqref{eq:minmod}].

\begin{figure}[t]
 \includegraphics[width=\columnwidth]{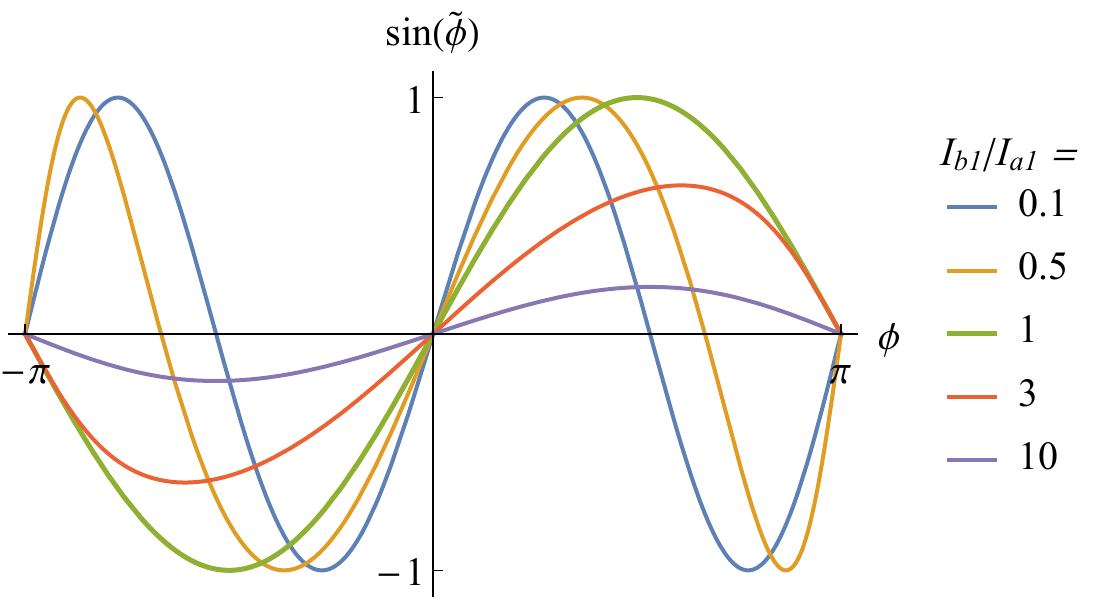}
\caption{Sine of the effective phase shift $\tilde\phi$ between the Josephson harmonics [see Eq.\ \eqref{eq:J}] as a function of the normalized SQUID flux $\phi=2\pi \Phi/\Phi_0$. This function determines the visibility of the JDE, and the JDE is absent if $\sin\tilde\phi=0$. This always takes place at $\phi=0 \bmod{\pi}$, and also at  intermediate values of $\phi$ in the case of $I_{b1}<I_{a1}$.}
 \label{fig:tildephi}
\end{figure}

The dependence of $\sin\tilde\phi$ on $\phi$ is illustrated in Fig.~\ref{fig:tildephi}. As long as $I_{b1}/I_{a1}<1$, the sine varies between $-1$ and $1$ crossing zero at ``symmetric'' points $\phi = 0 \bmod{\pi}$ and also at one additional (``accidental'') point between each two neighboring symmetric points. At $I_{b1}/I_{a1}=1$, we obtain $\tilde\phi = \phi$, and the accidental zeros disappear. Then, at $I_{b1}/I_{a1}>1$, the form of $\sin\tilde\phi$ is like a distorted $\sin\phi$ but with suppressed amplitude. The suppression is stronger at larger $I_{b1}/I_{a1}$.

The case of  $I_{b1}/I_{a1}>1$ is therefore less favorable from the point of view of observability of the diode effect. Experimentally, SQUIDs with  $I_{b1}/I_{a1}\leqslant 1$ are preferable.
Then why do we take into account the second Josephson harmonic $I_{b2}$ if $I_{b1}$ is not large? This may be justified if $I_{b1}$ is suppressed not due to interface transparency but due to additional physical mechanisms, e.g., near the $0$-$\pi$ transition \cite{Ryazanov2001,Kontos2002,Sellier2003,Sellier2004,Oboznov2006}. Note that the $0$-$\pi$ transition itself (sign change of $I_{b1}$) only gradually changes the $\tilde\phi(\phi)$ dependence, and so does not lead to qualitative changes in the diode effect.

\section{Conclusions}
\label{sec:conclusions}

In the framework of the RSJ model,
we have theoretically investigated asymmetric two-junction SQUIDs with different CPRs in the two JJs, involving higher Josephson harmonics ($\sin n\varphi$ with $n>1$).
We focused on the ``minimal model'' with one junction in the SQUID loop possessing purely sinusoidal current-phase relation (with $n=1$) and the other one featuring additional second harmonic (with $n=2$), see Eq.\ \eqref{eq:minmod}.
The (normalized) CPR of the SQUID then takes the form of Eq.\ \eqref{eq:J}. The effective phase shift $\tilde\phi$ between the Josephson harmonics depends on the magnetic flux through the interferometer loop.

Due to the presence of the second Josephson harmonic ($A\neq 0$), the CVC turns out to be asymmetric, $I(-V) \neq -I(V)$, in the cases when $\sin \tilde\phi \neq 0$.
The system thus demonstrates the JDE, the simplest manifestations of which is the direction dependence of the critical current, $I_c^+ \neq I_c^-$.
We employ a combination of perturbative methods to analytically describe asymmetry of the CVC, and then confirm our findings by numerical calculations in more general cases.
Efficiency of the JDE and its polarity are determined by $\tilde\phi$ and thus depend on the external magnetic flux.

We analytically demonstrate asymmetry of the overall $I(V)$ dependence in the limit of large current or small second harmonic in the absence of external ac irradiation.
In the presence of external irradiation of frequency $\omega$, phase locking of the external signal and the internal Josephson oscillations leads to peculiarities of the CVC at $\overline{V} =(n/k)\hbar\omega/2e$.

In the voltage-source case of external signal, the CVC demonstrates the Shapiro spikes [on the background of the linear ohmic $I(V)$ dependence]. The integer spikes (with integer $n/k$) are asymmetric, which is a manifestation of the JDE. The half-integer spikes (with half-integer $n/k$) remain symmetric.

In the current-source case, the CVC demonstrates the Shapiro steps. This case is more complicated for analytical consideration, and we consider it perturbatively with respect to the amplitude of the ac drive and of the second Josephson harmonic, and also numerically. Both integer and half-integer Shapiro steps are asymmetric, which is a manifestation of the JDE.

At present, systems of the considered type can be experimentally realized. We hope that our results will stimulate experimental research in this direction.

 \acknowledgments
We thank M.~V.\ Feigel'man, P.~M.\ Ostrovsky, and V.~V.\ Ryazanov for useful discussions.
The work was supported by the Russian Science Foundation (Grant No.\ 21-42-04410).
Ya.V.F.\ was also supported by
the Foundation for the Advancement of Theoretical Physics and Mathematics ``BASIS''.


\appendix

\section{Sinusoidal junction: details of the harmonic perturbation theory}
\label{sec:harmsindetails}

To formalize our perturbation theory, it is convenient to rescale time and frequency,
\begin{equation} \label{rescale}
\tau'=j\tau,\qquad \nu' = \nu/j.
\end{equation}
For brevity, we also introduce notations
\begin{equation}
C_n \equiv \cos n\nu' \tau' ,
\qquad
S_n \equiv \sin n\nu' \tau' .
\end{equation}

Instead of Eq.\ \eqref{varphi_eq}, we then have
\begin{equation} \label{Jeq1}
d\varphi / d\tau' = 1- \sin(\varphi) / j,
\end{equation}
and the form of solution is
\begin{equation} \label{phigen1}
\varphi(\tau') = \nu' \tau' + \sigma(\tau'),
\end{equation}
where
\begin{equation} \label{eqsigmatau}
\sigma(\tau') = \sum_{n=1}^\infty \left( a_n C_n + b_n S_n \right).
\end{equation}

The frequency and the coefficients in this expression are series with respect to $1/j$:
\begin{align}
\nu' &= \sum_{n=0}^{\infty} \nu'^{(n)},
\qquad
\nu'^{(n)} \sim 1 / j^n,
\\
a_n &= \sum_{k=0}^{\infty} a_n^{(k)} ,
\qquad
a_n^{(k)} \sim 1/ j^{n+k},
\label{ank} \\
b_n &= \sum_{n=0}^{\infty} b_n^{(k)},
\qquad
b_n^{(k)} \sim 1/ j^{n+k}
\label{bnk}
\end{align}
(some contributions may be zero, so the order-of-magnitude estimates refer only to those which are nonzero).
In each order of the perturbation theory, the sum of the indices of these quantities must be the same.
In the arguments of the cosines and sines, we write $\nu'$ as it is but as soon as it comes out to prefactors (due to taking the time derivative), we expand it to the relevant order.

The zeroth order is trivial: in Eq.\ \eqref{phigen1}, we leave only $\varphi = \nu'^{(0)} \tau'$, and Eq.\ \eqref{Jeq1} then gives
$\nu'^{(0)} = 1$.

\subsection{First order}

In the first order (with respect to $1/j$), we take into account the first harmonic from Eq.\ \eqref{phigen1}:
\begin{equation}
\varphi(\tau') = \nu' \tau'
+ a_1^{(0)} C_1 + b_1^{(0)} S_1,
\end{equation}
and the lhs of Eq.\ \eqref{Jeq1} in this order takes the form
\begin{equation} \label{lhs1}
d\varphi^{(1)} / d\tau' = \nu'^{(1)}
- a_1^{(0)} \nu'^{(0)} S_1 + b_1^{(0)} \nu'^{(0)} C_1.
\end{equation}
In the rhs of Eq.\ \eqref{Jeq1}, we must keep the terms of the first order:
\begin{equation} \label{rhs1}
\mathrm{rhs}^{(1)} =  -S_1/ j.
\end{equation}

Comparing the coefficients in front of
$S_n$ and $C_n$
in Eqs.\ \eqref{lhs1} and \eqref{rhs1},
we find
\begin{equation} \label{1order}
\nu'^{(1)} = 0,
\qquad
a_1^{(0)} = 1/j,
\qquad
b_1^{(0)} = 0.
\end{equation}

\subsection{Second order}

In the second order, we also take into account the second harmonic from Eq.\ \eqref{phigen1}:
\begin{widetext}
\begin{equation}
\varphi(\tau') = \nu' \tau'
+ \bigl( a_1^{(0)} + a_1^{(1)} \bigr) C_1 + \bigl( b_1^{(0)} + b_1^{(1)} \bigr) S_1
+ a_2^{(0)} C_2 + b_2^{(0)} S_2,
\end{equation}
and the lhs of Eq.\ \eqref{Jeq1} in this order takes the form
\begin{equation} \label{lhs2}
d\varphi^{(2)} / d\tau' = \nu'^{(2)}
- \bigl( a_1^{(0)} \nu'^{(1)} + a_1^{(1)} \nu'^{(0)} \bigr) S_1 + \bigl( b_1^{(0)} \nu'^{(1)} + b_1^{(1)} \nu'^{(0)} \bigr)  C_1
- 2 a_2^{(0)} \nu'^{(0)} S_2 + 2 b_2^{(0)} \nu'^{(0)} C_2.
\end{equation}

Now, for treating the second and higher orders, it is convenient to represent the rhs of Eq.\ \eqref{Jeq1} as
\begin{equation} \label{rhs}
\mathrm{rhs}
= 1- \sin(\nu' \tau' + \sigma) / j= 1- \left( S_1 \cos \sigma + C_1 \sin \sigma \right) /j.
\end{equation}
Below, we will expand $\cos \sigma$ and $\sin \sigma$ in Eq.\ \eqref{rhs} with respect to $\sigma$ (which is a small quantity) and then with respect to $a_n$ and $b_n$ in the required order of smallness.

The largest contribution to $\sigma$ is given by $a_1^{(0)}$ and $b_1^{(0)}$ which are of the order of $1/j$. Therefore, in the second order we represent Eq.\ \eqref{rhs} as
\begin{equation} \label{rhs2}
\mathrm{rhs}^{(2)} =  - \sigma^{(1)} C_1 /j = - \bigl( a_1^{(0)} C_1 + b_1^{(0)} S_1 \bigr) C_1 /j
= - \bigl[ a_1^{(0)} (1+C_2) + b_1^{(0)} S_2 \bigr] /2j.
\end{equation}

Comparing the coefficients in front of
$S_n$ and $C_n$
in Eqs.\ \eqref{lhs2} and \eqref{rhs2},
we find
\begin{equation}
\nu'^{(2)} = -1/ 2j^2,
\qquad
b_2^{(0)} = - 1 / 4j^2,
\qquad
a_1^{(1)} = b_1^{(1)} = a_2^{(0)} = 0.
\end{equation}

\subsection{Third order}

In the third order, taking into account that some lower-order coefficients are already found to be zero, we can write
\begin{equation}
\sigma =
\bigl( a_1^{(0)} + a_1^{(2)} \bigr) C_1 + b_1^{(2)} S_1
+ a_2^{(1)} C_2 + \bigl( b_2^{(0)} + b_2^{(1)} \bigr) S_2
+ a_3^{(0)} C_3 + b_3^{(0)} S_3,
\end{equation}
and the lhs of Eq.\ \eqref{Jeq1} in this order takes the form
\begin{equation} \label{lhs3}
d\varphi^{(3)} / d\tau'  = \nu'^{(3)}
+ \bigl( 1/2j^3- a_1^{(2)} \bigr) S_1 + b_1^{(2)} C_1
- 2 a_2^{(1)} S_2 + 2 b_2^{(1)} C_2
- 3 a_3^{(0)} S_3 + 3 b_3^{(0)} C_3.
\end{equation}

Now, in Eq.\ \eqref{rhs}, in $\cos \sigma$ and $\sin \sigma$ we must retain only the second-order contributions:
\begin{align}
\left( \cos \sigma \right)^{(2)} &= -\left( \sigma^2 \right)^{(2)} / 2 = - \bigl( a_1^{(0)} C_1\bigr)^2 /2 ,
\\
\left( \sin \sigma \right)^{(2)} &= \sigma^{(2)} = b_2^{(0)} S_2.
\end{align}
From Eq.\ \eqref{rhs} we then obtain
\begin{equation} \label{rhs3}
\mathrm{rhs}^{(3)}
= ( S_1 + S_3 ) / 4j^3 .
\end{equation}

Comparing the coefficients in front of
$S_n$ and $C_n$
in Eqs.\ \eqref{lhs3} and \eqref{rhs3}, we find
\begin{equation}
\nu'^{(3)} = 0,
\qquad
a_1^{(2)} = 1/ 4j^3,
\qquad
a_3^{(0)} = - 1/ 12 j^3,
\qquad
b_1^{(2)} = a_2^{(1)} = b_2^{(1)} = b_3^{(0)} = 0.
\end{equation}

\subsection{Fourth order}

In the fourth order, taking into account that some lower-order coefficients are already found to be zero, we can write
\begin{equation}
\sigma =
\bigl( a_1^{(0)} + a_1^{(2)}  + a_1^{(3)} \bigr) C_1 + b_1^{(3)} S_1
+ a_2^{(2)} C_2 + \bigl( b_2^{(0)} + b_2^{(2)} \bigr) S_2
+ \bigl( a_3^{(0)} + a_3^{(1)} \bigr) C_3 + b_3^{(1)} S_3
+ a_4^{(0)} C_4 + b_4^{(0)} S_4,
\end{equation}
and the lhs of Eq.\ \eqref{Jeq1} in this order takes the form
\begin{equation} \label{lhs4}
d\varphi^{(4)} / d\tau'
 = \nu'^{(4)}
- a_1^{(3)} S_1 + b_1^{(3)} C_1
- 2 a_2^{(2)} S_2 + 2  \bigl( 1 / 8j^4 + b_2^{(2)} \bigr) C_2
- 3 a_3^{(1)} S_3 + 3 b_3^{(1)} C_3
- 4 a_4^{(0)} S_4 + 4 b_4^{(0)}  C_4.
\end{equation}

Now, in Eq.\ \eqref{rhs}, in $\cos \sigma$ and $\sin \sigma$ we must retain only the third-order contributions:
\begin{align}
\left( \cos \sigma \right)^{(3)} &= - \left( \sigma^2 \right)^{(3)} / 2 = - a_1^{(0)} b_2^{(0)} C_1 S_2 ,
\\
\left( \sin \sigma \right)^{(3)} &= \left( \sigma - \sigma^3 /6 \right)^{(3)}
= a_1^{(2)} C_1
+ a_3^{(0)} C_3 - \bigl( a_1^{(0)} C_1 \bigr)^3 / 6.
\end{align}
From Eq.\ \eqref{rhs} we then obtain
\begin{equation} \label{rhs4}
\mathrm{rhs}^{(4)}
= -( 1 - C_4) / 8j^4.
\end{equation}

Comparing the coefficients in front of
$S_n$ and $C_n$
in Eqs.\ \eqref{lhs4} and \eqref{rhs4}, we find
\begin{equation}
\nu'^{(4)} = - 1 / 8 j^4,
\qquad
b_2^{(2)} =  - 1 / 8 j^4,
\qquad
b_4^{(0)} =  1 / 32 j^4,
\qquad
a_1^{(3)} = b_1^{(3)} = a_2^{(2)} = a_3^{(1)} = b_3^{(1)} =  a_4^{(0)} = 0.
\end{equation}

\subsection{Summary}

The algorithm formulated above can be continued up to any order.
What we have obtained considering the four orders, may be summarized in terms of frequency $\nu' = \nu'^{(0)}+ \nu'^{(2)}+ \nu'^{(4)}$. Rescaling back to the dimensionless frequency $\nu$ [see Eq.\ \eqref{rescale}], we obtain Eq.\ \eqref{eq:nuharm}.

\section{Asymmetric higher-harmonic SQUID: details of the harmonic perturbation theory}
\label{sec:harmnonsindetails}

The CPR for our model of asymmetric higher-harmonic SQUID is given by Eq.\ \eqref{eq:J}.
The RSJ equation \eqref{varphi_eq_J} then takes the form
\begin{equation} \label{eqwith2harm}
\dot \varphi = j - \sin\varphi - A \sin (2\varphi-\tilde\phi).
\end{equation}
We rescale time and frequency according to Eq.\ \eqref{rescale}, so that
\begin{equation} \label{Jeqasym}
d\varphi(\tau') / d\tau' = 1 - \bigl[ \sin\varphi + A \sin (2\varphi-\tilde\phi) \bigr] /j.
\end{equation}

We consider $|j|\gg 1$ and do not make any assumptions about $A$.
In the case of Eq.\ \eqref{Jeqasym}, the first and second harmonics in Eq.\ \eqref{eqsigmatau} should then be considered as being of the same order with respect to $1/j$ (since the second harmonics of solution are directly generated by the second Josephson harmonic in the CPR).
The zeroth order is $\nu'^{(0)} =1$.

\subsection{First order}

In the first order (with respect to $1/j$), we have
\begin{equation}
\varphi(\tau') = \nu' \tau' + \sigma^{(1)},
\qquad
\sigma^{(1)} =
a_1^{(0)} C_1 + b_1^{(0)} S_1
+ a_2^{(-1)} C_2 + b_2^{(-1)} S_2,
\end{equation}
where negative indices in the coefficients of the second harmonic are introduced in order to comply with the order-counting rules \eqref{ank}--\eqref{bnk}.

The lhs of Eq.\ \eqref{Jeqasym} in this order takes the form
\begin{equation} \label{lhsa}
d\varphi^{(1)} / d\tau' = \nu'^{(1)}
- a_1^{(0)} S_1 + b_1^{(0)} C_1
- 2 a_2^{(-1)} S_2 + 2 b_2^{(-1)} C_2.
\end{equation}
Writing the rhs of Eq.\ \eqref{Jeqasym} in the same order, we obtain simply
\begin{equation} \label{rhsa}
\mathrm{rhs}^{(1)} = - \bigl[ \sin\nu' \tau' + A \sin (2\nu' \tau'-\tilde\phi) \bigr] / j
= -\bigl( S_1 + A S_2 \cos \tilde\phi - A C_2 \sin\tilde\phi \bigr) / j .
\end{equation}
Comparing the coefficients in front of
$S_n$ and $C_n$
in Eqs.\ \eqref{lhsa} and \eqref{rhsa}, we find
\begin{equation}
\nu'^{(1)} = 0,
\qquad
a_1^{(0)} = 1/ j,
\qquad
b_1^{(0)} = 0,
\qquad
a_2^{(-1)} = A \cos (\tilde\phi) /2j,
\qquad
b_2^{(-1)} = A \sin (\tilde\phi)/ 2j.
\end{equation}

\subsection{Second order}

Now we consider the second order taking into account the third and the fourth harmonics in Eq.\ \eqref{eqsigmatau} as well as contributions of the same order to the coefficients of the first and the second harmonics:
\begin{gather}
\varphi(\tau')= \nu' \tau' + \sigma^{(1)} + \sigma^{(2)},
\qquad
\sigma^{(2)} = a_1^{(1)} C_1 + b_1^{(1)} S_1 +  a_2^{(0)} C_2 + b_2^{(0)} S_2 +  a_3^{(-1)} C_3 + b_3^{(-1)} S_3 + a_4^{(-2)} C_4 + b_4^{(-2)} S_4,
\\
d \varphi^{(2)} / d\tau'= \nu'^{(2)} - a_1^{(1)} S_1 + b_1^{(1)} C_1 - 2 a_2^{(0)} S_2 + 2 b_2^{(0)} C_2 - 3 a_3^{(-1)} S_3 + 3 b_3^{(-1)} C_3 - 4 a_4^{(-2)} S_4 + 4 b_4^{(-2)} C_4.
\label{eq:assymlhs2}
\end{gather}
Writing the rhs of Eq.\ \eqref{Jeqasym} in the same order, we obtain
\begin{multline}
\mathrm{rhs}^{(2)} = - \bigl[ \sigma^{(1)} \cos\nu' \tau' +2 A \sigma^{(1)} \cos (2\nu' \tau'-\tilde\phi) \bigr] / j
\\
= - \left[ (1+A^2) + C_2 + (5A/2) \bigl( C_1 \cos \tilde\phi  + S_1 \sin\tilde\phi  + C_3 \cos\tilde\phi  + S_3 \sin\tilde\phi \bigr)
+ A^2 \bigl( C_4 \cos 2\tilde\phi  + S_4 \sin 2\tilde\phi \bigr) \right] / 2j^2.
\label{eq:assymrhs2}
\end{multline}
Comparing the coefficients in front of
$S_n$ and $C_n$
in Eqs.\ \eqref{eq:assymlhs2} and \eqref{eq:assymrhs2}, we find
\begin{gather}
\nu'^{(2)} = -(1+A^2) / 2j^2,
\quad
a_1^{(1)} = 5A \sin(\tilde\phi)/4j^2,
\qquad
b_1^{(1)} = - 5A \cos(\tilde\phi)/ 4j^2,
\quad
a_2^{(0)} = 0,
\quad
b_2^{(0)} = - 1/ 4j^2,
\notag \\
a_3^{(-1)} = 5A \sin(\tilde\phi) / 12 j^2,
\quad
b_3^{(-1)} = - 5A \cos(\tilde\phi) / 12 j^2,
\quad
a_4^{(-2)} = A^2 \sin( 2\tilde\phi) / 8 j^2,
\quad
b_4^{(-2)} = - A^2 \cos( 2\tilde\phi) / 8 j^2.
\end{gather}

\subsection{Third order}

The third order leads to longer expressions, but we will be interested in finding $\nu'^{(3)}$ only. This quantity originates from the lhs of Eq.\ \eqref{Jeqasym}, while in the rhs of Eq.\ \eqref{Jeqasym} it is sufficient to keep $\sigma^{(1)}$ and $\sigma^{(2)}$ only. Indeed,
\begin{equation}
\mathrm{rhs}^{(3)} =
\Bigl[ \bigl( \sigma^{(1)} \bigr)^2 \sin \nu' \tau' - 2 \sigma^{(2)} \cos\nu' \tau'
+ 4 A \bigl( \sigma^{(1)} \bigr)^2 \sin(2\nu' \tau'-\tilde\phi) - 4 A \sigma^{(2)} \cos(2\nu' \tau'-\tilde\phi) \Bigr] / 2j.
\end{equation}
\end{widetext}
The constant (not containing $S_n$ and $C_n$) part of $\mathrm{rhs}^{(3)}$ is then easily found from this expression, and the result is
$\nu'^{(3)} = -3A \sin(\tilde\phi) / 4j^3$.

\subsection{Summary}

So, up to the third order we find
\begin{equation} \label{eq:omegalargej}
\nu' = 1-(1+A^2)/ 2j^2 - 3A \sin(\tilde\phi) /4j^3.
\end{equation}
Rolling back the rescaling of Eq.\ \eqref{rescale}, we find Eq.\ \eqref{eq:nulargej}.

\section{Direct integration method}
\label{sec:app:direct}

Alternatively to the method of Sec.~\ref{sec:alaThompsonPhi}, the perturbation theory with respect to the amplitude of the second Josephson harmonic can be constructed with the help of direct integration. Equation \eqref{eqstart1b} immediately yields
    \begin{equation} \label{eq:intdvarphi}
    \tau-\tau_0 =\int\frac{d\varphi}{j-\sin\varphi - A\sin( 2\varphi-\tilde\phi)},
    \end{equation}
and the integral in the rhs can, in principle, be calculated with the help of the universal trigonometric substitution $x = \tan(\varphi /2)$.
In order to implement this, we need to decompose the fraction in the rhs of Eq.\ \eqref{eq:intdvarphi}. At small $A$, the roots of the denominator can be found approximately.

In order to illustrate this method, below we consider the case of $\tilde\phi=0$.
For brevity, we put the integration constant $\tau_0$ equal to zero (which implies a shift of time origin).

\subsection{\texorpdfstring{$\tau(\varphi)$}{tau(phi)} and the CVC}

In the case of zero effective phase shift ($\tilde\phi=0$), we rewrite Eq.\ \eqref{eq:intdvarphi} as
    \begin{gather}
    \tau  = 2 L/j,
    \quad
    L= \int \frac{x^2+1}{P(x)} dx,
   \label{eq:tau-tau0} \\
    P(x)= (x^2+1) \left( x^2+1- 2x/j \right) +4A x(x^2-1)/j.
    \notag
    \end{gather}
The fourth-order polynomial $P(x)$ has two pairs of complex-conjugate roots: $a$, $a^*$, $b$, and $b^*$. At $A=0$, the two upper-half-plane roots are
    \begin{equation}
    a_0 =  i,
    \quad
    b_0 = \bigl(1+i\sqrt[\pm]{j^2-1} \bigr)/j.
    \end{equation}
So, $a$ in the limit $A\to 0$ becomes a ``trivial'' root, it is actually eliminated in this limit [due to the numerator under the integral in Eq.\ \eqref{eq:tau-tau0}], and the behavior of the sinusoidal junction is fully governed by the nontrivial $b$ root.

At nonzero $A$, we have
    \begin{equation}
      L=\int \frac{x^2+1}{(x-a)(x-a^*)(x-b)(x-b^*)} dx.
    \end{equation}
Expanding the integrand into elementary fractions, we find $L=L_{ab} + L_{ba}$, where
    \begin{equation}
      L_{ab} = \frac 1{\Im a} \Im\left[ \frac{(a^2+1) \ln(x-a)}{(a-b)(a-b^*)} \right] ,
        \label{eq:Iab}
    \end{equation}
and $L_{ba}$ is obtained from $L_{ab}$ after interchanging $a\leftrightarrow b$.

Our variable $x=\tan(\varphi/2)$ sweeps the whole real axis when $\varphi$ changes from $-\pi$ to $\pi$. Further increase of $\varphi$ must be accompanied by a branch switching of the logarithm [that enters Eq.\ \eqref{eq:Iab}], so that the $L(\varphi)$ function grows on average.

In order to establish the correct branch choice, we can make a step back to the purely sinusoidal case ($A=0$). In this case, $L_{ab}$ (which corresponds to the $a_0$ poles) vanishes, and only the $L_{ba}$ contributes to $L$, which then takes the form
    \begin{equation}
      L_0 = \Im \left[ \ln(x-b_0) \right] / \Im b_0.
    \end{equation}
Considering the logarithm in the complex plane with the $(-\infty,0]$
cut, so that
\begin{equation}
\arg z = -\arctan ( \Re z / \Im z ) + (\pi / 2) \sgn(\Im z),
\end{equation}
we find
    \begin{equation} \label{eq:I0}
      L_0 = \frac 1{\Im b_0} \left( \arctan \frac{x-\Re b_0}{\Im b_0} - \frac\pi 2 \right),
    \end{equation}
which finally yields
    \begin{equation} \label{eq:tau-tau0unp}
      \tau  = \frac 2{\sqrt[\pm]{j^2-1}} \left( \pi n_\varphi + \arctan \frac{j \tan \varphi/ 2 -1}{\sqrt[\pm]{j^2-1}} \right),
    \end{equation}
    with $n_\varphi = \left\lfloor \varphi/2\pi \right\rceil$
[note that we have eliminated the last $\pi/2$ term in Eq.\ \eqref{eq:I0} since it can be absorbed into the integration constant $\tau_0$]. Of course, this result reproduces Eq.\ \eqref{exactsol2}.

Returning to the  case of nonzero $A$,
we calculate $L$ taking into account the branch switching of the $\arg$ function. As a result, instead of Eq.\ \eqref{eq:tau-tau0unp} we now find
    \begin{align}
      \tau
      &= A_{ab} \biggl( \pi n_\varphi + \arctan \frac{\tan \varphi /2 -\Re a}{\Im a} \biggr)
     \notag \\
      + & A_{ba} \biggl( \! \pi n_\varphi + \arctan \frac{\tan\varphi /2 -\Re b}{\Im b} \! \biggr)
      \! + \! B \ln \left| \frac{\tan\varphi /2-a}{\tan\varphi /2-b} \right| \! ,
    \label{tau(phi)}
    \end{align}
where
\begin{equation}
      A_{ab} = \Re C,
      \quad
      B = \Im C,
      \quad C = \frac{2(a^2+1)}{j\Im (a) (a-b)(a-b^*)},
\end{equation}
and $A_{ba}$ is obtained from $A_{ab}$ after interchanging $a\leftrightarrow b$.

At $n\to \infty$, we thus have the asymptotic behavior $\varphi \simeq 2\pi n_\varphi$ (by definition of $n_\varphi$) and $\tau \simeq (A_{ab}+A_{ba}) \pi n_\varphi$ [from Eq.\  \eqref{tau(phi)}],
hence we find the average voltage bias
    \begin{equation} \label{eq:nudirectint}
      \overline{v} = \lim_{\tau\to\infty} \varphi(\tau)/\tau = 2 / (A_{ab}+A_{ba}) .
    \end{equation}

The results of Eqs.\ \eqref{tau(phi)} and \eqref{eq:nudirectint} are general in the sense that the second-harmonic amplitude $A$ can be arbitrary.
We have expressed the result in terms of the roots of the $P(x)$ polynomial.

Now, we consider the case of small $A$.
Up to the second order with respect to $A$, the roots of $P(x)$ are
\begin{align}
a &= i + 2i A + A^2 (2i-4j),
\\
b &= (1-2A) \bigl( 1+i\sqrt[\pm]{j^2-1} \bigr) /j + A^2 j\bigl( 4- 2i / \sqrt[\pm]{j^2-1} \bigr),
\notag
\end{align}
hence
    \begin{align}
      A_{ab} &= 24 j A^2, \qquad
      B =-4A,
      \\
      A_{ba} &= 2/\sqrt[\pm]{j^2-1} - 4(6j^4-9j^2+2) A^2 / \sqrt[\pm]{j^2-1}^3.
      \notag
    \end{align}
Equation \eqref{eq:nudirectint}
then yields
    \begin{multline} \label{eq:vdirect}
      \overline{v} =\sqrt[\pm]{j^2-1}- 2 A^2 \biggl[ 6j(j^2-1) - \frac{6j^4-9j^2+2}{\sqrt[\pm]{j^2-1}} \biggr] .
    \end{multline}
To test this result, we can consider the limit $|j|\gg 1$, in which case Eq.\ \eqref{eq:vdirect} reproduces the $\tilde\phi=0$ limit of Eq.\ \eqref{eq:nulargej} in the order $1/j^2$ (up to which the comparison is possible).

\subsection{\texorpdfstring{$\varphi(\tau)$}{phi(tau)}}

As we have seen in the previous subsection, knowledge of the (inverse) $\tau(\varphi)$ dependence is sufficient for studying the CVC in the case of fixed current.
At the same time, inverting the dependence given by Eq.\ \eqref{tau(phi)}  in order to obtain the direct $\varphi(\tau)$ function (which is required, e.g., for calculating the Shapiro steps) cannot be performed in the general case.

Nevertheless, this inverting can be done perturbatively in the case of small $A$.
Below we illustrate this procedure in the first order with respect to $A$.
We write Eq.\ \eqref{tau(phi)} as an expansion with respect to $A$:
\begin{equation} \label{eq:taufvarphi}
  \tau = g_0(\varphi) + A g_1(\varphi).
\end{equation}
We want to invert Eq.\ \eqref{eq:taufvarphi} as
\begin{equation} \label{eq:f0Af1}
  \varphi = f_0(\tau) + A f_1(\tau) .
\end{equation}
In the zeroth order, $g_0(\varphi)$ is given by Eq.\ \eqref{eq:tau-tau0unp}, and its inverse $f_0(\tau)$
is given by Eq.\ \eqref{exactsol2}.

Substituting Eq.\ \eqref{eq:f0Af1} into Eq.\ \eqref{eq:taufvarphi} and expanding up to the first order, we obtain
\begin{equation}
  \tau = g_0(f_0(\tau)) + g_0'(f_0(\tau)) A f_1(\tau) + A g_1(f_0(\tau)) .
\end{equation}
Considering the terms of the first order, we find
\begin{equation}
  f_1(\tau)  = - g_1 / g_0' ,
  \label{eq:g1}\\
\end{equation}
where the derivative is taken with respect to $\varphi$, and finally $f_0(\tau)$ is substituted as the argument in the rhs.
Since
  $g_0'(\varphi) = 1/(j-\sin\varphi)$,
we find
\begin{equation}
  g_0'(f_0(\tau)) = (1+\cos\alpha \cos\theta(\tau)) \cos\alpha / \sin^2 \alpha ,
\end{equation}
where  $\theta (\tau) = \nu \tau - \alpha$.

From Eq.\ \eqref{tau(phi)}, we find
\begin{equation}
  g_1(\varphi) = -2 \left[ \ln (j g_0') +1 - j g_0' \right],
\end{equation}
and Eq.\ \eqref{eq:g1} finally leads to the following explicit result of the inversion procedure:
\begin{multline}
  f_1(\tau) = \frac{2}{1+\cos\alpha \cos\theta(\tau)} \biggl[ - \cos\alpha - \cos\theta(\tau)
  \\
  + \frac{\sin^2 \alpha}{\cos\alpha} \ln\left( \frac{1+\cos\alpha \cos\theta(\tau)}{\sin^2 \alpha} \right)\biggr].
\end{multline}
Note that this result reproduces Eq.\ \eqref{eq:varphi1Thompson} in the case of $\tilde\phi=0$ if we choose the free constant $\tau_1$ in Eq.\ \eqref{eq:varphi1Thompson} according to $\cos\theta(\tau_1) = -\cos\alpha$.

\section{Shapiro spikes: details of derivation}
\label{sec:app:spikes}

Derivation of equations describing the Shapiro spikes (see Sec.~\ref{sec:spikes}) is based on the following relation:
\begin{equation} \label{eq:sinphitophi}
e^{iz\sin\varphi} = \sum_{n=-\infty}^\infty J_n(z) e^{i n\varphi} ,
\end{equation}
where  $J_n(z)$ are the Bessel functions of the first kind. A useful property of these functions is $J_{-n} (z) = (-1)^n J_n(z)$. Multiplying both sides of Eq.\ \eqref{eq:sinphitophi} by $e^{ia}$ and taking the imaginary part (in the case of real $a$ and $z$), we find
\begin{align}
\sin(a + z\sin\varphi)
&= \sum_{n=-\infty}^\infty J_n(z) \sin(a+ n\varphi)
\notag \\
&= \sum_{n=-\infty}^\infty (-1)^n J_n(z) \sin(a- n\varphi).
\end{align}
This leads to Eq.\ \eqref{eq:Jkn}.

\input{Josephson_diode_PRB.bbl}

\end{document}

%% file: Josephson_diode_PRB.bbl
%